\documentclass[pre,aps]{revtex4-2}
\usepackage{amsfonts}
\usepackage{times}
\usepackage{amsmath}
\usepackage{graphicx}
\usepackage{amssymb}
\usepackage{xcolor,comment}

\newcommand{\bk}{\mathbf{k}}
\newcommand{\br}{\mathbf{r}}

\newcommand{\bI}{\mathbf{I}}

\begin{document}

\title{Solitary  wave billiards}

\author{Jes\'us Cuevas-Maraver}
\affiliation{Grupo de F\'{\i}sica No Lineal, Departamento de F\'{\i}sica Aplicada I,
Universidad de Sevilla. Escuela Polit\'{e}cnica Superior, C/ Virgen de Africa, 7, 41011-Sevilla, Spain}
\affiliation{Instituto de Matem\'{a}ticas de la Universidad de Sevilla (IMUS). Edificio
Celestino Mutis. Avda. Reina Mercedes s/n, 41012-Sevilla, Spain, Avda Reina Mercedes s/n, 41012 Sevilla, Spain}

\author{Panayotis G.\ Kevrekidis}
\affiliation{Department of Mathematics and Statistics, University
of Massachusetts, Amherst, Massachusetts 01003-4515, USA}

\author{Hong-Kun Zhang}
\affiliation{Department of Mathematics and Statistics, University
of Massachusetts, Amherst, Massachusetts 01003-4515, USA}

\begin{abstract}
In the present work we { explore} the concept of solitary wave
billiards. I.e., instead of a point particle, we
examine a
solitary wave in an enclosed region and explore its collision
with the boundaries and the resulting trajectories in cases which
for particle billiards are known to be integrable and for
cases that are known to be chaotic. A principal conclusion
is that solitary wave billiards are generically found to
be chaotic even in cases where the classical particle billiards are
integrable. However, the degree of resulting chaoticity depends
on the particle speed and on the properties of the potential. Furthermore, the nature of the
scattering of the deformable solitary wave
particle is elucidated on the basis of a
negative Goos-H{\"a}nchen effect which, in
addition to a trajectory shift, also results
in an effective shrinkage of the billiard domain.
\end{abstract}

\maketitle

\section{Introduction}

In recent years, the study of solitary waves has been a topic
of widespread appeal in a broad range of fields. For instance,
relevant coherent structures arise in the exploration of
electrical field dynamics within optical fibers~\cite{hasegawa,agrawal},
as well as in the study of nonlinear effects in plasmas~\cite{plasmabook}.
They emerge in atomic clouds of Bose-Einstein condensates (BECs)~\cite{pethick,stringari}, as well as in so-called rogue
waves in the ocean~\cite{slunaev}. Of central role within most of
the above studies and fields has been the envelope wave model of the
nonlinear Schr{\"o}dinger equation~\cite{ablowitz,ablowitz2,ablowitz1,sulem},
which can be used to describe the electric field of light, the
wavefunction of BEC atoms, or the water wave elevation. Indeed,
the resulting nonlinearity may stem from different sources, such
as e.g., the so-called Kerr effect in optics~\cite{hasegawa,agrawal}
or an effective mean-field nonlinearity due to contact interactions
between bosons~\cite{pethick,stringari}, yet its impact is similar
across fields in producing robust solitary waves of either
bright~\cite{sulem,agrawal}, dark~\cite{siambook} or rogue~\cite{slunaev}
form.

On the other hand, the notion of billiards is one that
has attracted significant attention both at the classical,
as well as at the quantum side. Here, a point particle
is reflected within a (typically) enclosed 2D domain $Q$ (of different possible
shapes)~\cite{sinai1,chernov} (see also~\cite{mathworld}
for an extensive set of references). The particle moves
freely aside from its interaction with the boundary $\partial Q$,
by elastic reflections without
loss of speed. Depending on the nature of the billiard
(e.g., square or circular or instead an elliptical
stadium~\cite{bunim1,bunim2} or a
square with an enclosed circle~\cite{sinai1,sinai2}), the
outcome of the reflections could be a closed trajectory
associated with integrable motion, or an ergodic
trajectory associated with a chaotic billiard.

While the above aspects involve classical billiards,
we do note in passing the consideration of quantum billiards.
The quantum mechanical billiard is given by the linear Schr{\"o}dinger equation typically with Dirichlet boundary condition. It describes a wave function associated with a probability density $|\psi|^2$ within
the domain while the wall features an infinite potential wall.
A relevant summary of some of the activity in the context of
experiments (and theory) in the theme of microwave
billiards and the connections to quantum chaos can be
found in~\cite{scholar}, where a number of references
to this theme are summarized.
The dynamics of quantum billiard  are determined by the Hamiltonian equations of motion on the domain $Q$. The wave function $\psi$ of a single quantum particle with unit mass obeys the Schr{\"o}dinger equation
\begin{equation}\label{LinearSch} i\frac{\partial \psi}{\partial t}=-\nabla^2 \psi+V(\br)\psi,\,\,\,\,\,\,\,\,\,\,\,\, \br=(x,y)\in int(Q)
\end{equation}
with, e.g., Dirichlet boundary conditions  $\psi(x,y)=0$, for $(x,y)\in \partial Q$.
  Usually one can assume in such a setting that $V(x,y)=0$ for $(x,y)\in \text{int}( Q)$, and $V(x,y)=\infty$, for $(x,y)\in \partial Q$.  Most research has been concentrated, for  such quantum billiards, on the
  search for quantum chaos. Even though it was proved that if classical billiard is ergodic, then the corresponding quantum billiard is quantum ergodic, so-called quantum ``scars" are observed in most cases, 
  which demonstrate that quantum billiards are more complicated.
  The Bunimovich
  stadium~\cite{bunim1,bunim2}  has recently become a  very popular quantum billiard model, while the study of quantum chaos
  in different billiards is generally a topic that is gaining
  considerable traction~\cite{proz0,proz1,proz2}.

  A less studied  billiard is the ``soft" version, where the boundary of the billiard table $Q$ is defined  by a finite potential wall $V$, with $V(x,y)=0$, for $(x,y)\in \text{int}(Q)$, and $V(x,y)\in (0,\infty)$ for $(x,y)\in \partial Q$.
  In such a setting, for a Sinai (dispersing) billiard~\cite{sinai1,sinai2},   the existence of a stable island around a periodic orbit which is tangent to the billiard’s wall (or near the corner) was found numerically in~\cite{KaplanFriedman};
  see also~\cite{Kroetz2016,Gil-Gallegos2019} for other examples of
  such soft billiards.

  In the present work our aim is to {
  explore} a notion that in a
  sense interweaves the above concepts while paving
  { an avenue}
  involving the interplay of nonlinearity, billiards and possibly
  quantum mechanics (but also perhaps more concretely in our examples
  below, nonlinear
  optics) in the form of {\it solitary wave billiards}.
  In particular,
  the connections are quite natural: solitary waves are
  typically thought of as effective particles (indeed, hence the
  name)~\cite{agrawal,ablowitz,ablowitz2,ablowitz1,sulem,siambook}.
  Therefore, it is natural to examine these effective particles
  in a billiard setting. At the same time, nonlinear Schr{\"o}dinger
  equation models appear in a natural sense to generalize the concept
  of linear quantum billiards in the presence of different types of nonlinearity,
  stemming from nonlinear optics~\cite{agrawal}, plasmas~\cite{plasmabook} or atomic BECs~\cite{siambook}. In that light too, it is useful to adapt
  some of the ideas of linear quantum billiards to such more complex
  cases where the localized wavefunction is held together because
  of the interplay of nonlinearity and dispersion while slowly
  losing some of its initial kinetic energy through the (potential) interactions
  with the wall. In that vein, the notion of a soft billiard is a
  natural one as the solitary wave interacts with a finite potential wall.
  However, the interaction may be inelastic even in the context of an
  infinite potential due to the non-integrability of the PDE model.
  The latter is the typical scenario in the associated $2+1$-dimensional
  spatio-temporal models. It is worthwhile to mention here that the
  idea of billiards in settings such as BECs has not only been
  theoretically explored~\cite{zhang,shep}, but also experimentally
  realized in the form of an optically induced confinement potential
  (i.e., billiard) for the ultracold atoms~\cite{raizen}.

  However, { exploring} such solitary wave billiard ideas requires
  some extra care. One needs to avoid the hurdle of collapse
  (arising, e.g., in $2+1$-dimensional cubic nonlinearity models~\cite{sulem}).
  Instead one needs to consider a nonlinearity that allows for a robust
  $2+1$-dimensional standing wave; here the saturable nonlinearity
  of relevance to photorefractive crystals is thus selected~\cite{jianke}
  as a physically relevant example.
  { While in such systems, pattern formation has been much
    studied earlier~\cite{arrecchi}, here we have in mind a different
    setting well summarized in the authoritative review of~\cite{dso}.
    In particular, as is explained therein [see, e.g., Fig. 3.4 and associated discussion],
 optical beams of interfering plane waves in the ordinary polarization cause
 an effective crystal lattice potential
 for the extraordinary polarization. Rather, our proposal (which,
while not realized to our knowledge, seems well within the reach of
such experiments) is to utilize
suitable beams in the context of the ordinary polarization in order to form, e.g.,
the billiard configurations proposed below}.
       Then, the solitary wave can
       be boosted using a Galilean transformation
  and the billiard dynamics will be accordingly initiated. In what
  follows, we discuss our numerical observations and an emerging
  qualitative understanding of the relevant phenomenology.

  We also
  note in passing that, to our knowledge, no Hamiltonian variant of this
  type has been considered in the literature. A dissipative (but
  { occurring over
  a narrow region of the associated
  parameter space})
  analogue of such concepts has been presented
  in~\cite{Lugiato}; {
  see also~\cite{prati}}.
  { A pioneering
  effort in a related direction involved
  the experimental demonstration
 of the interaction of the
  so-called walkers (a droplet
 bouncing on vibrating bath) and their
  interactions with barriers,
  as shown, e.g., in the work
  of~\cite{eddi}. Indeed, the associated
field of hydrodynamic quantum analogs
has seen a wide range of developments
that have now been summarized in
reviews such as~\cite{john}.}

  { As we will see below, the present
    setting
    will offer a family of (stationary) solitary waves in terms of
    their amplitude/width
    as we will showcase below, but also the freedom to boost these
    solitary waves at essentially arbitrary speeds based on the above
    transformation.
    We consider these features to be key to the potential interest in the
    setting
    that we propose herein.}
  Our presentation of the results is structured as follows. In section II,
  we present the mathematical setup of the problem. In section III,
  we discuss the numerical observations and their comparison to the
  point billiard model. Finally, in section IV, we summarize our
  findings and present some conclusions and directions for future study.

\section{Mathematical Setup}

\subsection{Soliton wave billiard} We start from a two-dimensional nonlinear Schr\"odinger (NLS) equation with (saturable) photorefractive nonlinearity~\cite{jianke},
\begin{equation}
\mathrm{i}\psi_t=\mathcal{L}(\psi),\,\,\,\,\,\,\mathcal{L}(\psi)=-\nabla^2\psi+V(\br)\psi-\frac{2|\psi|^2}{1+|\psi|^2}\psi,\,\,\,\,\,\,\,\,\br=(x,y)\in Q
\label{eq:dyn}
\end{equation}
with $V(\br)$ being the potential where the relevant geometry of the billiard will be introduced on $Q$, and $\nabla^2\psi=\psi_{xx}+\psi_{yy}$. Here, $\psi(\br,t)$
represents the envelope of the electromagnetic field for the optical problem of interest, with the density $\rho=|\psi|^2$
corresponding to the light intensity. While the particular nonlinearity
is less directly relevant to atomic BECs, one can envision other
nonlinearities there (e.g., competing ones as in the recently
budding field of quantum droplets~\cite{boris}) that can lead
to a similar phenomenology. In the latter case $\psi(\br,t)$
plays the role of the condensate wavefunction, while $|\psi|^2$
represents the atomic density.
In the optical setting, the potential $V(\br)$ stems from the
refractive index profile~\cite{kivshar}, while in BECs, the ability
to ``paint'' arbitrary (and, hence, also billiard) potentials has been
demonstrated~\cite{boshier}.

The above dynamical equation can be derived from
a Hamiltonian formulation as:
\begin{equation}
i\psi_t=\frac{\delta H[\psi]}{\delta \psi^*}
\end{equation}
where $H[\psi]$ is the (conserved during the dynamics)
Hamiltonian functional:
\begin{equation}
H[\psi]=\int_{\br\in Q} \left[|\nabla\psi|^2+\left(V(\br)-2\right)|\psi|^2+2\log(1+|\psi|^2)\right]\,\mathrm{d}\br.
\label{eq:energy}
\end{equation}

The initial condition for our dynamical simulations will be a stationary solitary wave set into motion by means of a Galilean boost.
This is a desirable feature of the present model
as the traveling waves
herein
can be generically achieved and with arbitrarily selected speeds due
to the Galilean invariance.
The stationary state is found by introducing the ansatz
\begin{equation}
    \psi(x,y;t)=\exp(\mathrm{i}\omega t)\phi_{\omega}(x,y)
\end{equation}
into Eq.~(\ref{eq:dyn}), and a fixed point algorithm with finite-differences discretization has been used to this aim.
It follows that $\phi_{\omega}$, if it exists, is a nonlinear  eigenfunction of $\mathcal{F}$:
\begin{equation}
-\omega \phi_{\omega}=\mathcal{F} (\phi_{\omega}) =-\nabla^2\phi_{\omega}+V(x,y)\phi_{\omega}-\frac{2|\phi_{\omega}|^2}{1+|\phi_{\omega}|^2}\phi_{\omega}.
\label{ss_eqn}
\end{equation}

Once this stationary wave is found, we fix $\omega$, and the initial condition of the simulation is attained from the Galilean boost of the stationary structure centered at $(x,y)=(0,0)$.
\begin{equation}\label{psi0}
    \psi_0(x,y)\equiv\psi(x,y;0)=\phi_{\omega}(x,y)\exp(i (k_xx+k_yy))=\phi_{\omega}(x,y)\exp(i k(x\cos\theta+y\sin\theta))
\end{equation}
leading to a wave moving with an initial velocity $v_0=2|\bk|$ that forms an angle $\theta$ with the $x$-axis, and
\begin{equation}
\bk=(k_x, k_y)=|\bk|(\cos\theta, \sin\theta)=k (\cos\theta, \sin\theta).\label{v0}
\end{equation}
Consequently, the energy imparted to the moving coherent structure (in comparison to its stationary state) is given by
\begin{equation}
    \Delta E\equiv E[\psi_0(x,y)]-E[\phi_{\omega}(x,y)]=\int \mathrm{d}^2\!r\left[|\nabla\psi_0|^2-|\nabla\phi_{\omega}|^2\right]=N |\mathbf{k}|^2
\end{equation}
 and is naturally associated with the excess kinetic energy, with $N$ being the
 squared $L^2$ norm or the mass, as it is often referred to,
 of the stationary state
\begin{equation}
    N=\int_Q |\phi_{\omega}(\br)|^2\,d\br.
\end{equation}
Assigning $N \equiv 2m$ to an effective mass,
we can express this
excess energy as:
\begin{equation}\label{eq:kinetic}
    \Delta E=2m|\mathbf{k}|^2=\frac{1}{2}mv_0^2
\end{equation}

If we suppose that the solitary is moving in a vanishing potential landscape,
it can be considered as a quasi-particle that freely moves
with a initial kinetic energy equal to $\Delta E$, and this  quasi-particle
can be interpreted as possessing an effective mass $m$.

Note that at each time $t\geq 0$, the  solitonic wave function $\psi(x,y,t)$, which starts from $\psi_0(x,y)$, defines a reference probability measure $\nu(t)$ on the $\sigma$-algebra of the billiard table $Q$,  with density function $$\rho(\br,t)=\frac{|\psi(\br,t)|^2}{\int |\psi(\br,t)|^2 \, d\br}$$
Using this reference measure, we can investigate some physical  quantities (given by expected position, variance, and more general moments)
of the solitary wave billiard.
To study this billiard $\psi(x,y,t)$, we  first consider its center of
mass trajectory $\gamma(t)$, which describes the expected position of
the coherent structure. More precisely,   the expected position of the
localized pattern with respect to the reference measure $\nu$ is
defined as $$\gamma(t):=\mathbb{E}_{\nu_t}(\br)=\int_Q (x,y)
\rho(\br,t)\, d\br.$$

Since $\br=(x,y)$, we  can also write $\gamma(t)=(X(t), Y(t))$ with
$$X(t)=\mathbb{E}_{\nu_t}(x)=\int x\,d\nu_t,\,\,\,\, Y(t)= \mathbb{E}_{\nu_t}(y)=\int y\,d\nu_t, $$as the expected $x$ and $y$ position of the solitary wave, respectively.
Note that $(X(t),Y(t))$ can also be interpreted as the center position of the relevant pattern. In the following sections, we will mainly focus on the study of $\br(t)=(X(t), Y(t))$ numerically.

To assess how well the expected position  $\bar\br(t)$ of the solitary wave within the billiard describes its motion, we also calculate the variance of $\br$ with respect to the reference measure $\nu(t)$:
$$\text{Var}_{\nu_t}(\br):=\mathbb{E}_{\nu_t}(\|\br-\gamma(t)\|^2)=\int_{\br\in Q}
    \left((x-X(t))^2+(y-Y(t))^2)\right)\rho(\br,t) d\br$$
This quantity characterizes the width of our solitary wave.
For narrower solitary waves, it is enough to study the center of mass orbit $\gamma(t)$,  corresponding to the solitonic wavefunction $\psi(x,y,t)$.

\section{Numerical Results}

\subsection{Numerical simulation setup of the dynamical orbit
within the solitary wave billiard}

In our simulations, to simplify the calculations, we first locate the solitary wave at the point of maximal probability density $\rho(\br,t)$, which is denoted as  $\tilde \gamma(t)=(\tilde{x}(t),\tilde{y}(t))$ such that for any $t\geq 0$,
$$\rho(\tilde\gamma(t))=\max_{\br\in Q}\rho( \br,t)$$
Our computations suggest that the curve $\rho(\tilde\gamma(t))$ is well-defined, i.e. for each $t$, there is only one (principal, i.e.,
above a selected cutoff) local maximum for the density function.
We subsequently consider a window of size $2\delta\times 2\delta$ centered on $\tilde \gamma(t)$, and denote it as $U_{\delta}(t)\subset Q$;
this is suitably adjusted when the wave is in the vicinity of
the boundary.
In our simulation, we select $\delta=4$.

The conditional probability measure $\hat\nu_t:=\nu|_{U_{\delta}(t)}$ with density function $\hat\rho$ can be calculated as $$\hat\rho(\br,t)=\frac{|\psi(\br,t)|^2\,\bI_{U_{\delta}(t)}}{\int_{U_{\delta}(t)} |\psi(\br,t)|^2 \, d\br}$$

Moreover,  the simulated expected position of the solitary wave
within the billiard with respect to the conditional measure $\hat\nu$ can be calculated as
$$(\hat X(t),\hat Y(t)):=\mathbb{E}_{\hat\nu_t}(\br)=\int_Q (x,y) \hat\rho(\br,t)\, d\br.$$
This also provides a smoother and more accurate evaluation of the center of mass, in
comparison to the argmax over the numerical grid.
Furthermore, the variance associated with the solitary wave
position is given by:
$$W(t):=\int_{\br\in Q}
    \left((x-\hat X(t))^2+(y-\hat Y(t))^2)\right)\hat \rho(\br,t) d\br$$

\begin{figure}[tbp]
\begin{tabular}{cc}
\includegraphics[width=9cm]{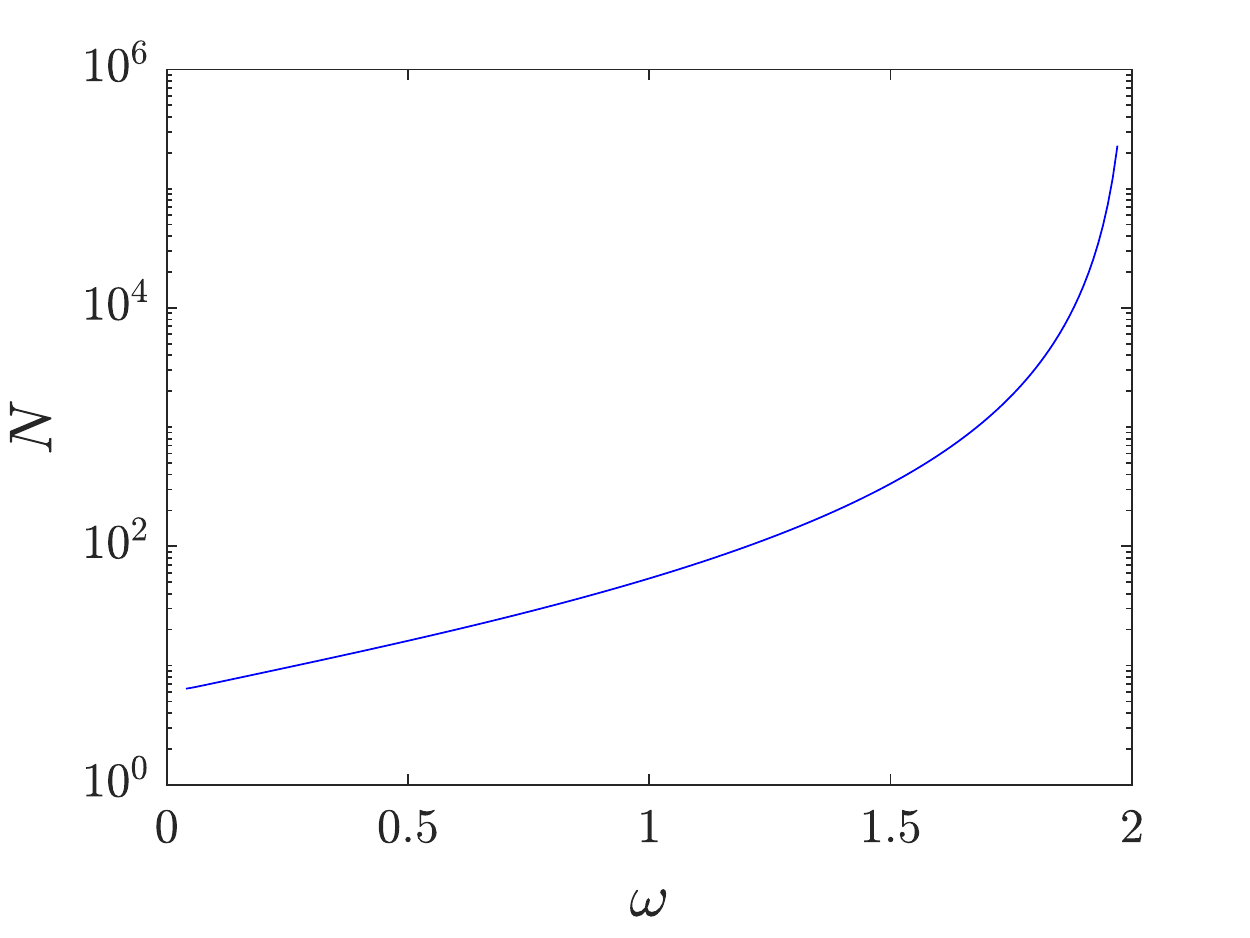} &
\includegraphics[width=9cm]{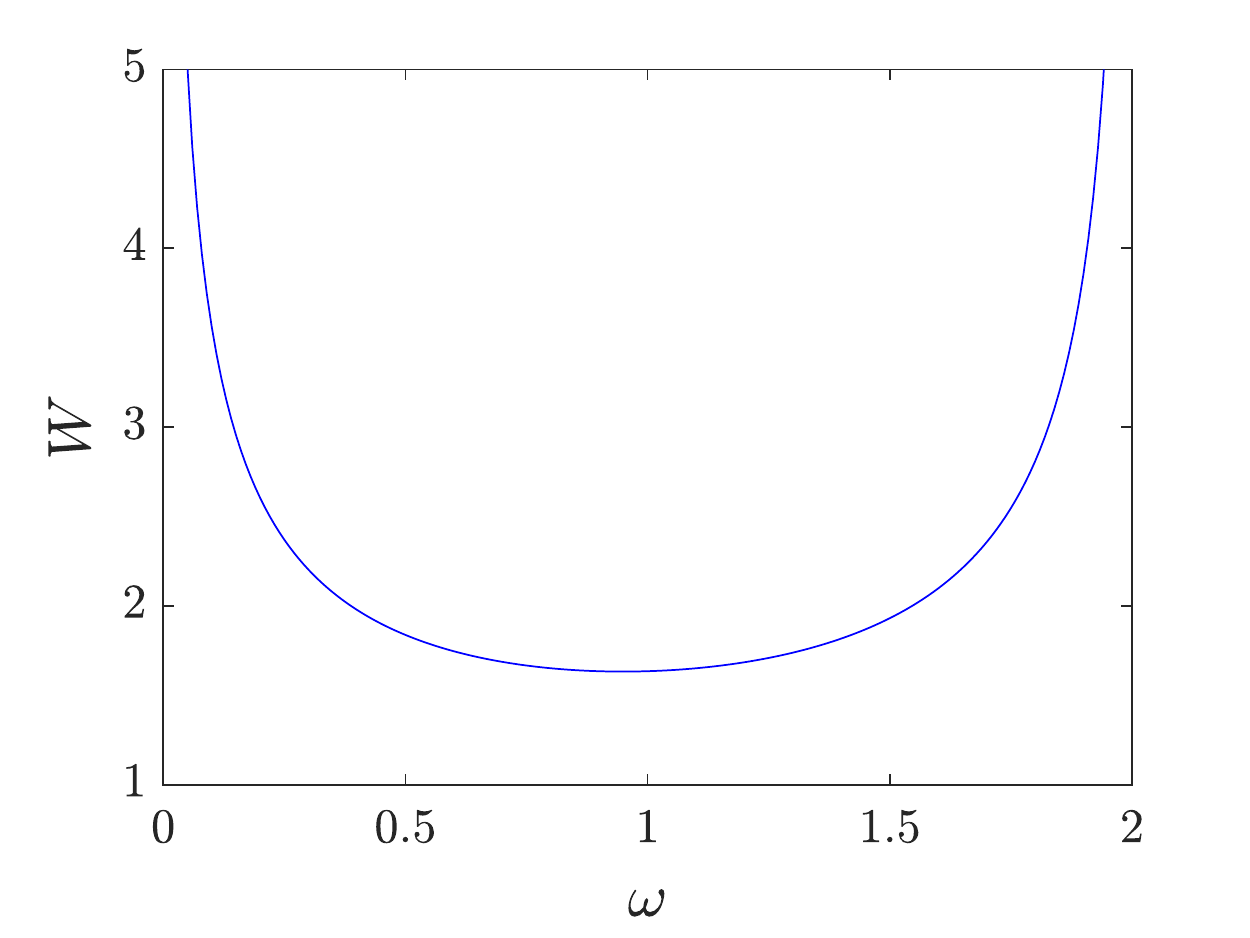} \\
\end{tabular}%
\caption{Dependence of the norm and the width
of the solitary wave with respect to its frequency
for the relevant family of solutions.}
\label{fig:norm}
\end{figure}

In Fig.~\ref{fig:norm},
 one can see that the narrowest solitary wave is found for $\omega=0.95$, and that the associated frequency lies in $\omega\in(0,2)$ interval, with the lower limit corresponding to the bifurcation of the solitary from the linear modes (continuous spectrum) band, while the upper limit reflects
 a divergence of its mass. {Indeed, the continuous spectrum of the problem
 lies in the interval $(-\infty,0)$, hence the avoidance of resonances therewith
 leads to solitary waves with $\omega>0$. On the other hand, the large intensity
 limit of the nonlinearity leads the nonlinear term to the asymptotic value of $-2$,
 hence Eq.~(\ref{ss_eqn}) accordingly becomes (in the large intensity limit)
 $(\omega-2) \phi_{\omega}=\nabla^2 \phi_{\omega}$, which requires
 $\omega<2$. As a result, solutions are expected to be found for
 $0<\omega<2$ in line with our numerical results.
 }

In our study, we mainly consider waves with  frequency $\omega=0.5$.
In the case examples that follow, we will either consider
infinite potentials (implemented via Dirichlet boundary conditions)
or finite potentials that are strictly higher than the above-mentioned
kinetic
energy of the solitary wave, so as to ensure a genuine billiard situation in which
the coherent structure will be reflected from the potential walls.

\subsection{Goos-H{\"a}nchen shifts of reflection upon potential walls in billiards of constant potential}

Before delving into the systematics of the solitary wave
trajectories within the entire billiard,
we briefly
highlight a key feature of the  billiard reflection of
the wave upon collision with the potential walls, for the case when the potential $V$ is finite at the boundary of $Q$.

When describing total  reflection of a solitary wave in such
a billiard potential setting, we need to identify expressions for the phase shift that occurs for each such event between the incident and reflected waves as a function of angle of incidence.
{A way  to describe the phase shift is as being due} to the orbit actually travelling a small distance into the potential wall before being reflected. An incident orbit behaves as if it were laterally displaced upon reflection. This feature has been previously recognized
as being analogous to the so-called Goos-H{\"a}nchen shift in optics,
occurring also to solitary waves during their interaction
with an interface; see, e.g., Ref.~\cite{curto} and the
discussion therein.
We will discuss this in further quantitative detail
in what follows below.
It is interesting to note
here that the Goos-H{\"a}nchen effect (GHE)
has been recently encountered not only in the
context of solitary waves interacting with
external potentials but also of linear waves
interacting with solitary ones~\cite{magnon}.

We first consider a solitary wave
placed in a billiard involving a square potential.  It is known that classical square billiards are completely  integrable~\cite{arithm}. Moreover, the orbits are periodic if  $\tan\theta$ is rational, for the collision angle $\theta$.
When studying the evolution of solitary waves, we have firstly considered a square barrier potential of amplitude $\alpha$, i.e.,
\begin{equation}
    V(x,y)=
    \begin{cases}
    \alpha & \quad\text{if } |x|>3L/4 \text{ or } |y|>3L/4 \\
    0 & \quad\text{otherwise}
    \end{cases}
    \label{eq:square_step}
\end{equation}
 The domain is taken with $L=40$.
This potential is shown on the left panel of Fig.~\ref{fig:square_potential}. {The evolution of dynamical equations has been found by means of numerical integration using a fixed step $\delta t=0.0625$ 4th-order Runge--Kutta algorithm with a spatial discretization $\delta x=0.5$. The latter value is reasonable enough as the soliton width is several times that value. In fact, we have used smaller values as 0.2 or 0.1 and have corroborated that the results do not change in any observable manner
(although the latter cases utilize far larger computational resources and hence
are considerably slower, hence the selection of $\delta x$ used for our
presentation herein). {{ We have also checked that, for this choice of $\delta t$ and $\delta x$, the relative error for the norm, i.e. $|N(t)/N(0)-1|$ does not
grow above $10^{-6}$ for the time horizons
considered.}

\begin{figure}[tbp]
\begin{tabular}{cc}
\includegraphics[width=9cm]{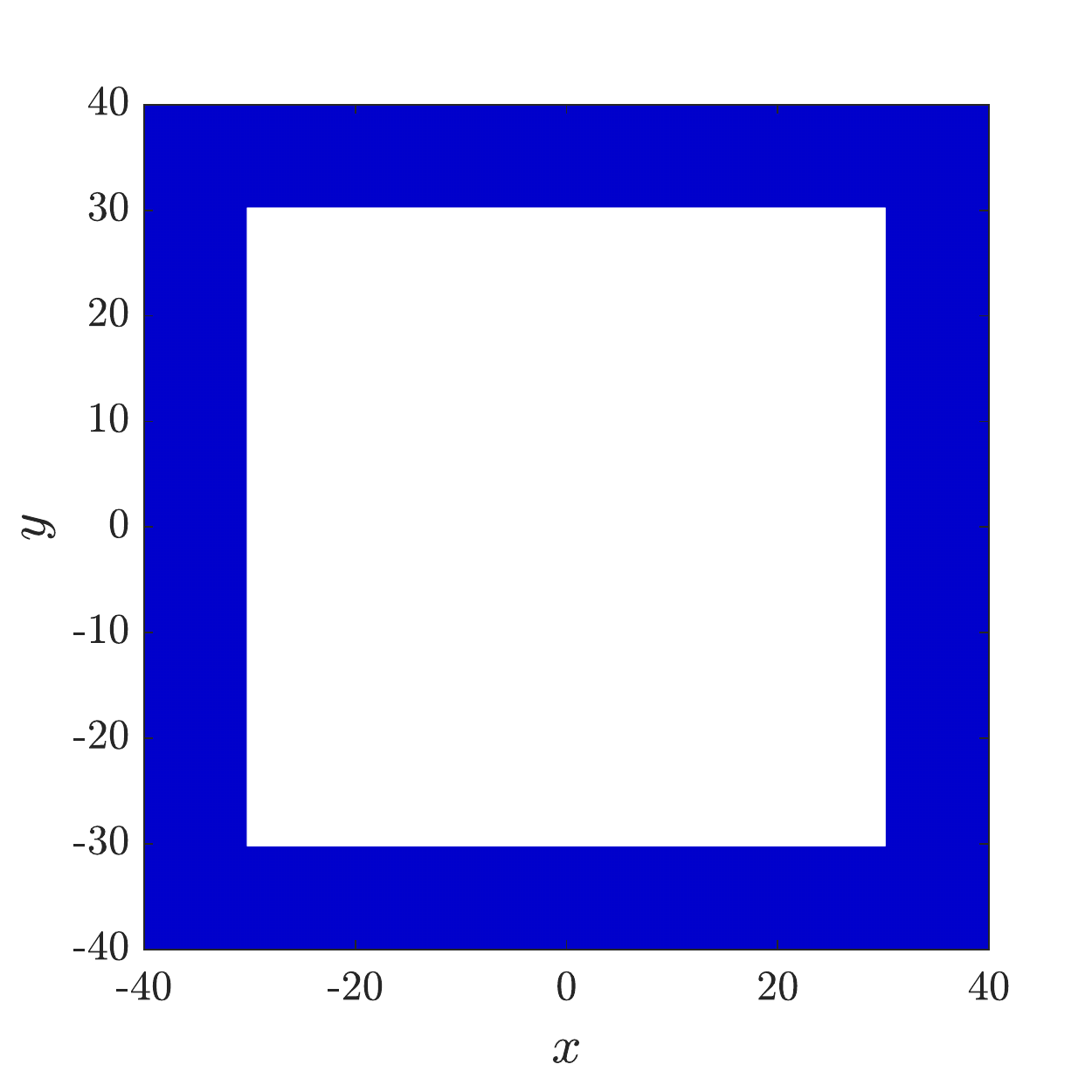} &
\includegraphics[width=9cm]{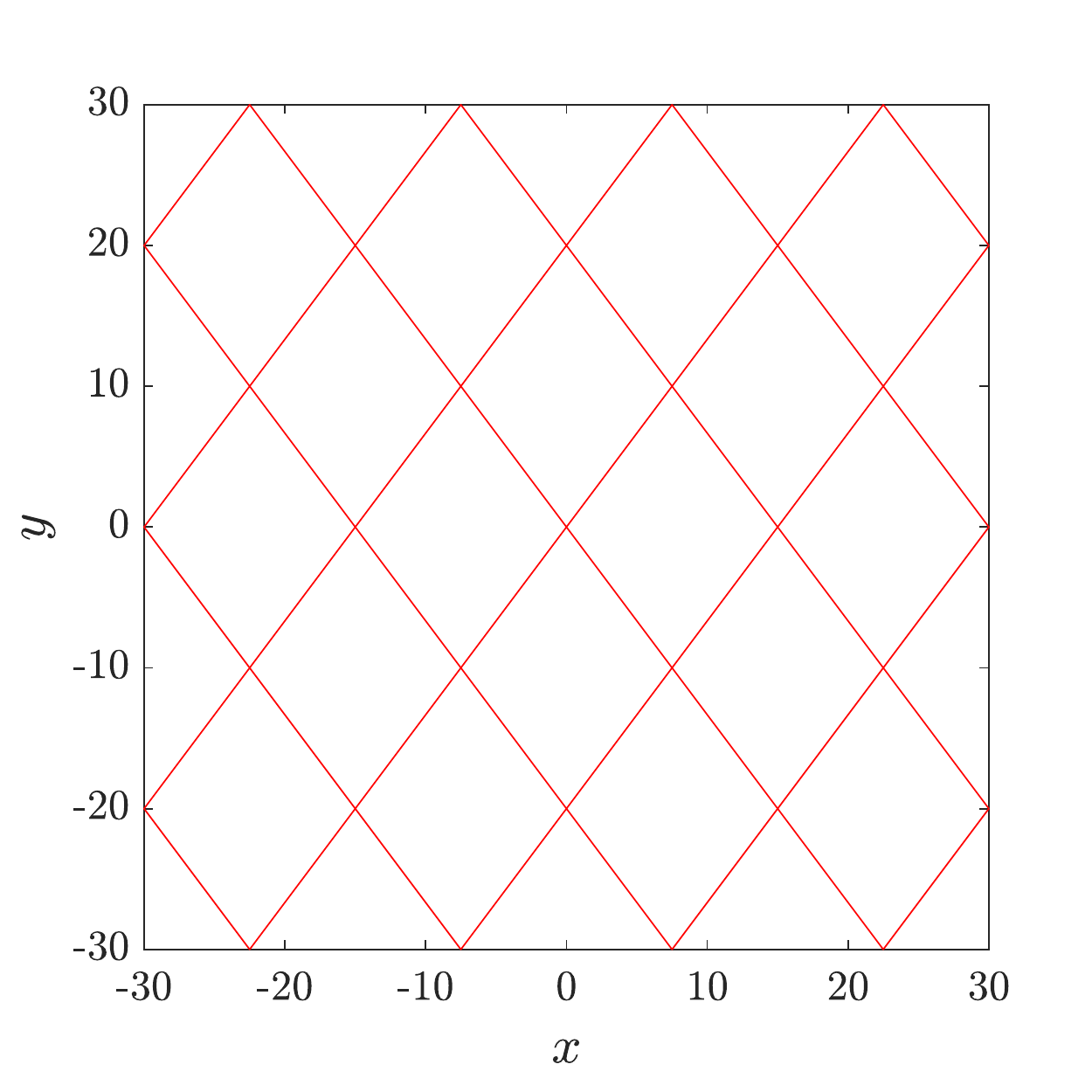} \\
\end{tabular}%
\caption{{(Left) Form of the square potential barrier given by (\ref{eq:square_step}). (Right) The trajectory of a classical {\it point} particle (right), in this potential.}}
\label{fig:square_potential}
\end{figure}

{In what follows, we have chosen  $\cos\theta=3/5$ for the initial collision angle. The right panel of Fig.~\ref{fig:square_potential}, depicts the path of a classical point particle
in this square potential. One can indeed see,
in line with the theoretical expectation, that the dynamics is periodic for this choice of $\theta$.}
On the other hand, in the left panel of Fig.~\ref{fig:square1}, the path of the solitary
wave with $v_0=0.02$ and $\alpha=4$ is considered. One can see that {contrary to the classical billiard case, the dynamics is not periodic. In addition,} the finite size of the solitary wave
particle does not allow its center to reach
the edge of the potential barrier (at $\pm 3L/4$
in each direction), but rather it effectively
gets reflected ``sooner''.
This is caused by the  reflection rule of the trajectory of the solitary wave upon a constant potential (similar to what is observed in \cite{Lugiato}).
Nevertheless, it is important to recognize that the solitary wave
 is not an infinitesimal particle mass. Instead, it features a
 finite width and for that reason this effective particle is
 reflected while its center of mass is still at a considerable
 distance from the domain boundary. There is some tunability
 of the width of the solitary wave (controlled by $\omega$), however,
 the constraint of avoiding the cubic nonlinearity and associated
 collapse phenomenology does not allow for the freedom
 of (effectively) arbitrarily selecting the solitonic width,
 as, e.g., can be done in one-dimensional such systems. Rather,
 here we are practically considering not only a soft potential
 (unless we examine the Dirichlet boundary condition case), but
 also a non-vanishing size particle that features a finite width and
 can, in principle, manifest internal modes capable of storing
 energy. This ``deformability'' of the solitary wave can also
 be detected in Fig.~\ref{fig:square1} at the positions of
 wall interactions where the trajectory of the solitary wave
 curves around in the reflection process, a manifestation
 of the GHE discussed above (see also details below).

As an interesting additional case example along this vein, in the right panel of Fig.~\ref{fig:square1}, we have considered the example of
the structure with the smallest width, namely $\omega=0.95$. 
Interestingly, as we can see in that case the trajectory does not close, indicating the subtle nature of the corresponding
phenomenology. As highlighted above, it is no longer the case that the dynamical evolution purely depends on the angle of incidence, but also on other features of the soft particle, such as its width, its potential excitation of internal modes (or dispersive radiation modes, as will
be seen below), etc. Hence, potential closed orbits should be considered to be a non-generic scenario for the deformable, solitary wave particles
of interest herein. This can also be confirmed in the consideration of settings with different $\alpha$ (such as $\alpha=10$) or different boundary conditions (such as Dirichlet),{{ for which, as we have checked, the soliton follows a very similar trajectory to that of the $\alpha=4$ case.}

\begin{figure}[tbp]
\begin{tabular}{cc}
\includegraphics[width=9cm]{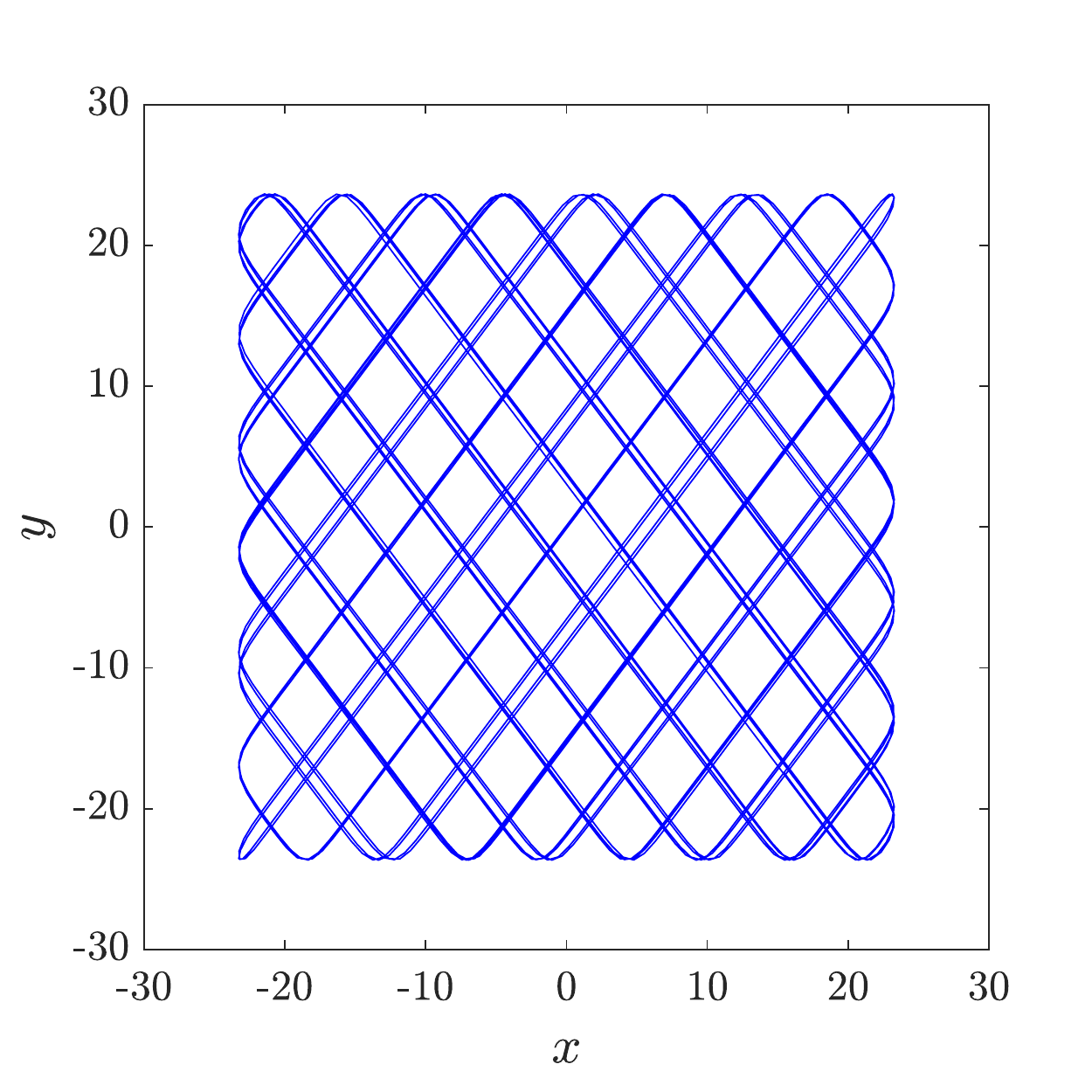} &
\includegraphics[width=9cm]{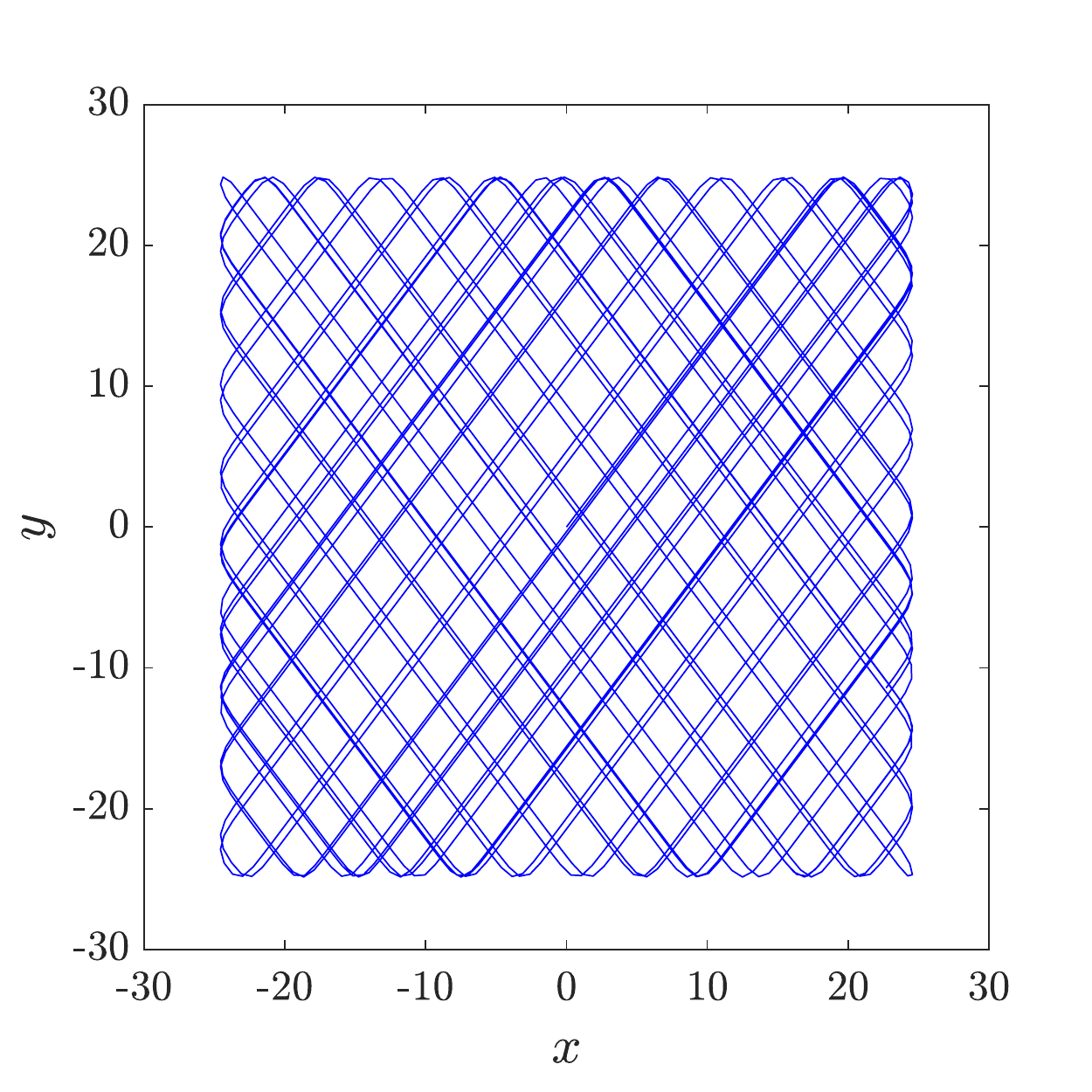} \\
\end{tabular}%
\caption{{Evolution of the center-of-mass of the finite-width solitary with $\alpha=4$ and $v_0=0.02$, for $\omega=0.5$ (left) and $\omega=0.95$ (right). The final time of both simulations is $t=2\times10^5$. Notice that in this figure and the other ones showing square barriers, the axis limits corresponds to the barrier boundaries. See also companion movies \texttt{movie1.gif} (corresponding to $\omega=0.5$) and \texttt{movie2.gif} (corresponding to $\omega=0.95$) in \cite{movies} depicting the evolution of the solitary wave dynamics, {in which the whole domain is shown with the barrier limits being depicted as white lines.}}}
\label{fig:square1}
\end{figure}

We have also explored how the above dynamics is altered when the velocity is increased and the barrier height is kept fixed; see, e.g., Fig.~\ref{fig:square2} for the solitary wave with $v_0=0.1$ and $\alpha=4$. The resulting motion in this case as well is not periodic. If the wave moves more rapidly ($v_0=0.8$), it deforms nontrivially in the collisions with the barrier, and some radiation is emitted.
The evolution of the coherent structure motion can be observed in more detail in the companion movies \cite{movies}, where the  density ($|\psi(x,y;t)|^2$) and its logarithm is tracked. {One can see how the solitary wave approaches the barrier and
is reflected after each collision. This process is central to our observation of the trajectories of the solitary waves (and their non-closed form), hence we focus on this further in what follows. However, before doing so, we point out the substantial difference between this high speed case example and the ones shown earlier, as concerns the substantial emission of dispersive wave radiation. This results in the solitary wave
moving within a ``turbulent'' small amplitude background, which, in turn, weakly affects (via scattering processes) the solitary wave motion. While this aspect is not pursued further herein, we find it an interesting direction for further exploration in the future. { Let us remark that in this particular case,
given the faster growth of the error
in the 4th order Runge-Kutta integrator,
we have numerically integrated the evolution by means of an adaptive step-size Dormand-Prince algorithm keeping the relative error below $10^{-7}$ during the whole integration time.}

\begin{figure}[tbp]
\begin{tabular}{cc}
\includegraphics[width=9cm]{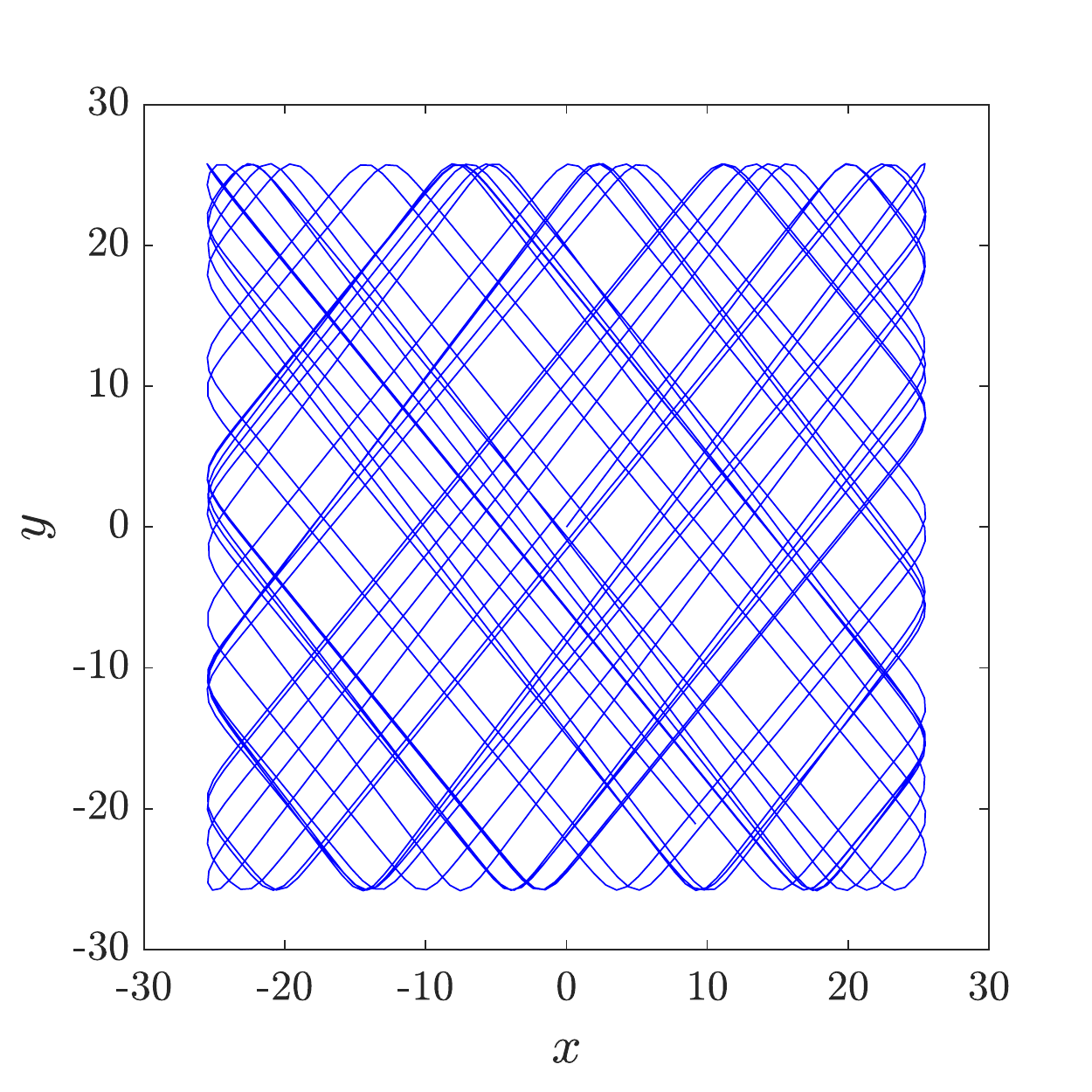} &
\includegraphics[width=9cm]{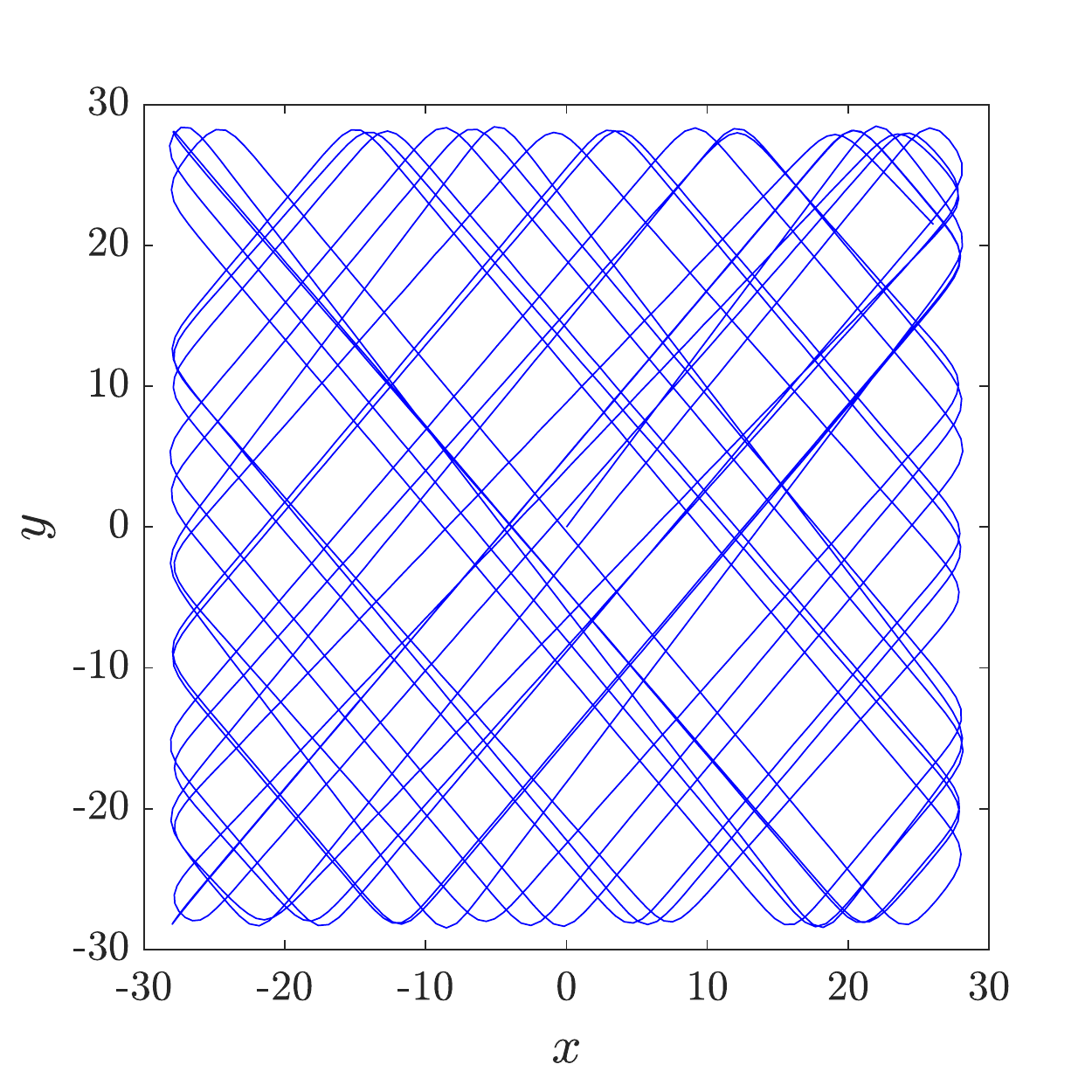} \\
\end{tabular}%
\caption{Evolution of the center-of-mass of the solitary waves with  $v_0=0.1$ (left panel) and $v_0=0.8$ (right panel), in the square barrier potential (\ref{eq:square_step}). In both cases, $\alpha=4$ and $\omega=0.5$. The final times of the simulations are $t=4\times10^4$ and $t=5\times10^3$, respectively. See also companion movies \texttt{movie3.gif} and \texttt{movie4.gif} in \cite{movies} depicting the evolution of the solitary wave dynamics.}
\label{fig:square2}
\end{figure}

{An explanation for the fact that the wave's center of mass does not reach the barrier is based on the intriguing
observation of a {\it negative} Goos-H{\"a}nchen effect (see e.g. \cite{Rechts} for such an effect in one-dimensional optical systems, as well as~\cite{curto} for
a solitonic example). Figure~\ref{fig:GHE1} shows the evolution of the integrated norm with respect to $x$-coordinate, i.e. $|\psi_x(y,t)|^2=\int \mathrm{d}x |\psi(x,y,t)|^2$ together with a scheme defining the Goos-H{\"a}nchen shift (GHs), $\Delta$, for the collision of the solitary
wave with a wall barrier of height $\alpha$.}
{As is well known, the GHE occurs when a beam of
finite extent is incident on a medium with smaller refraction index
with an angle higher than the critical angle for total reflection
(recall that $-V(x,y)$ plays the role of refraction index detuning in
optics). The shift is caused by the field penetrating the medium for a
small distance, forming a non-uniform plane wave that is evanescent in
the direction normal to the interface and propagating also along
it. The shift is positive in the usual case of plane waves flowing
parallel to the interface in the forward direction. There are,
however, cases when the  wave
reflects from a point in front of the contact
point in the {geometric optics} picture; in these cases, observed
for plane waves at interfaces with metals and predicted for
metamaterials, the plane waves flow parallel to the interface in the
backward direction and the GHE is negative; in addition, part of the
beam can cross the interface despite the negative GHE, as is the case
in metamaterials. All the aforementioned behaviour holds for linear
waves. In the case of nonlinear waves, in order to provide an analogy
with geometric optics, one can trace straight lines tangent to the
trajectory of the soliton's center of mass at the initial point, in a
fashion similar to what is done in Fig.~\ref{fig:GHE1}. }

\begin{figure}[tbp]
\begin{tabular}{cc}
\includegraphics[width=9cm]{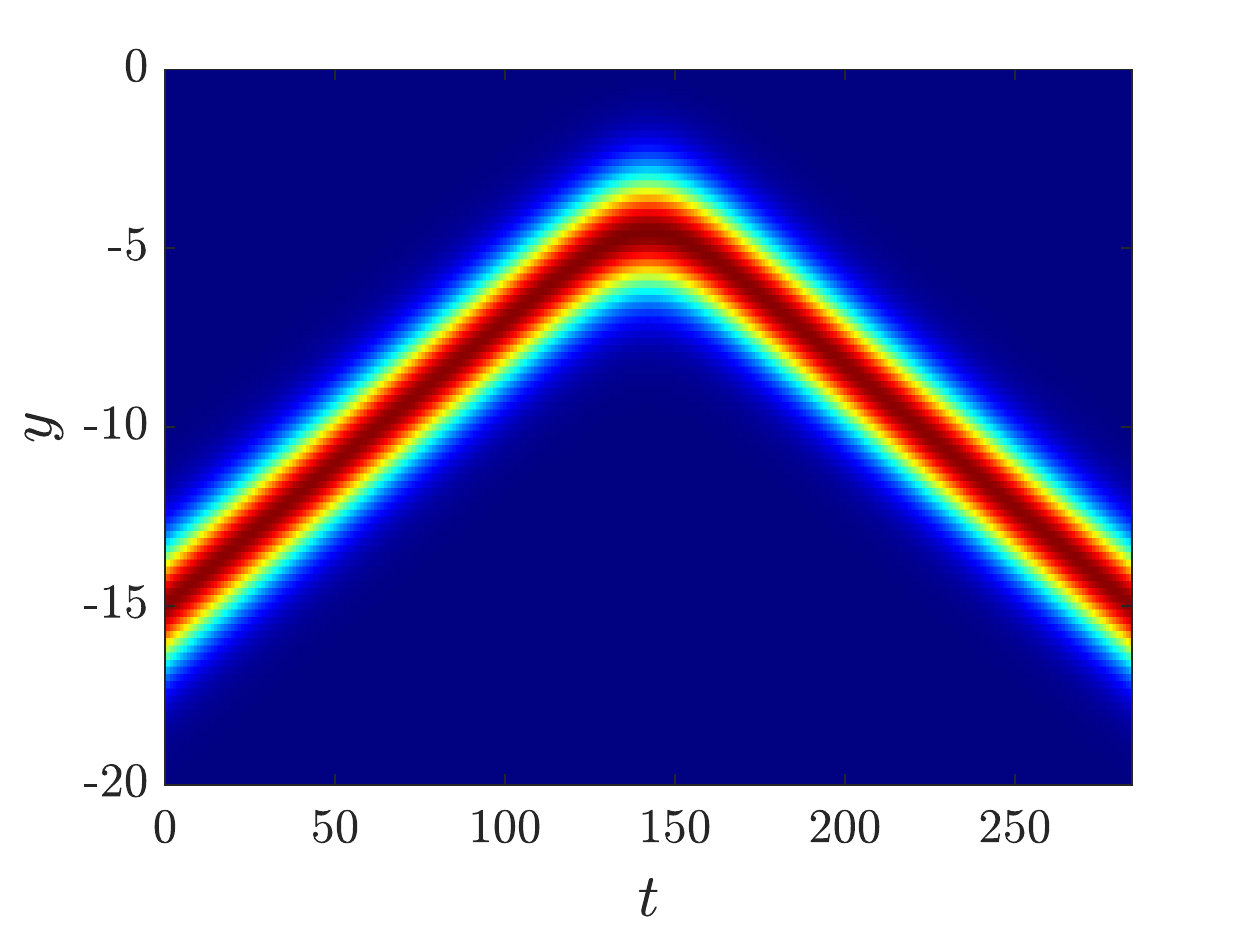} &
\includegraphics[width=9cm]{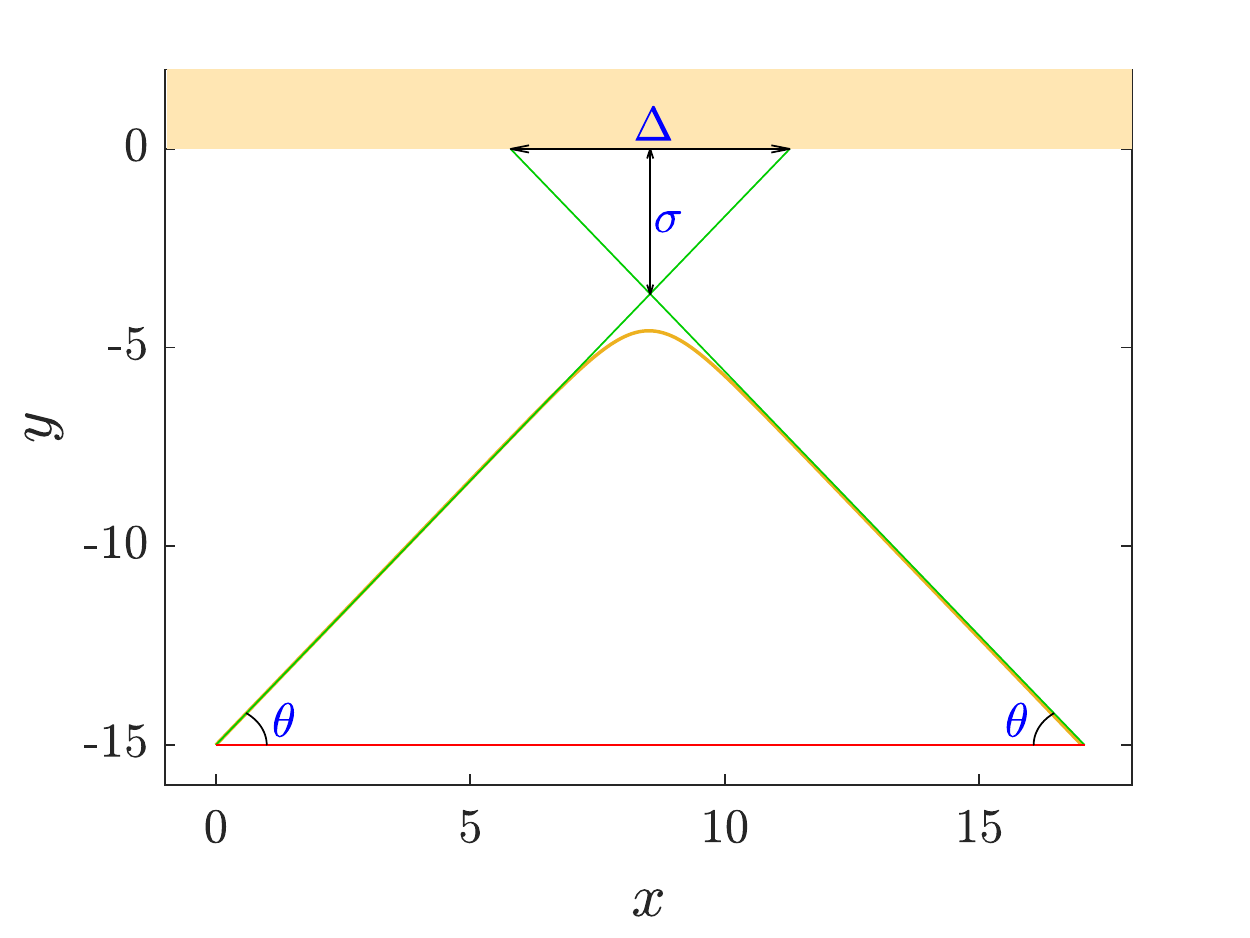} \\
\end{tabular}%
\caption{(Left panel) Evolution of the projected solitary wave density when interacting with a wall barrier of height $\alpha=4$ located at $y=0$; the initial velocity of the coherent structure is $v_0=0.1$ and the initial angle $\theta=\cos^{-1}3/5$. (Right panel) Path followed by the center of mass of the wave together with a
scheme indicating some relevant parameters like the
Goos-H\"anchen shift $\Delta$ or the domain shrinking $\sigma$.}
\label{fig:GHE1}
\end{figure}

{Notice that because of the GHs, the motion of the wave's center of mass when interacting with the wall can be approximated by a classical particle in a domain shrunk by a distance $\sigma=(\Delta\cot\theta)/2$. An important point is that in the square barrier, the impact angle of the solitary wave with the vertical lines of the square is the complementary to the angle of impact with the horizontal lines. Consequently, there will be two different GH shifts to account for, namely the one corresponding to an initial angle $\theta$ (denoted as $\Delta$), and the one for $\theta_c=\pi/2-\theta$ (denoted as $\Delta_c$, with the corresponding shrinkage $\sigma_c=(\Delta_c\tan\theta)/2$) in the perpendicular direction. Then, the trajectory of the center of mass of the solitary wave can be approximated by a classical particle in a rectangle bounded by $|x|=3L/4-\sigma_c$ and $|y|=3L/4-\sigma$.}
{Figure~\ref{fig:GHE2} shows the numerically
computed dependence of the domain shrinking caused by the GHE with respect to $v_0$ for fixed $\omega=0.5$, and also the dependence with $\omega$ for $v_0=0.1$. One can see that in the latter case, the minimum values of $\sigma$ and $\sigma_c$ are not attained for the narrowest solitary wave, i.e., that with $\omega=0.95$, but rather around $\omega=1.05$ (the particular value depends on the barrier height).}

\begin{figure}[tbp]
\begin{tabular}{cc}
\includegraphics[width=9cm]{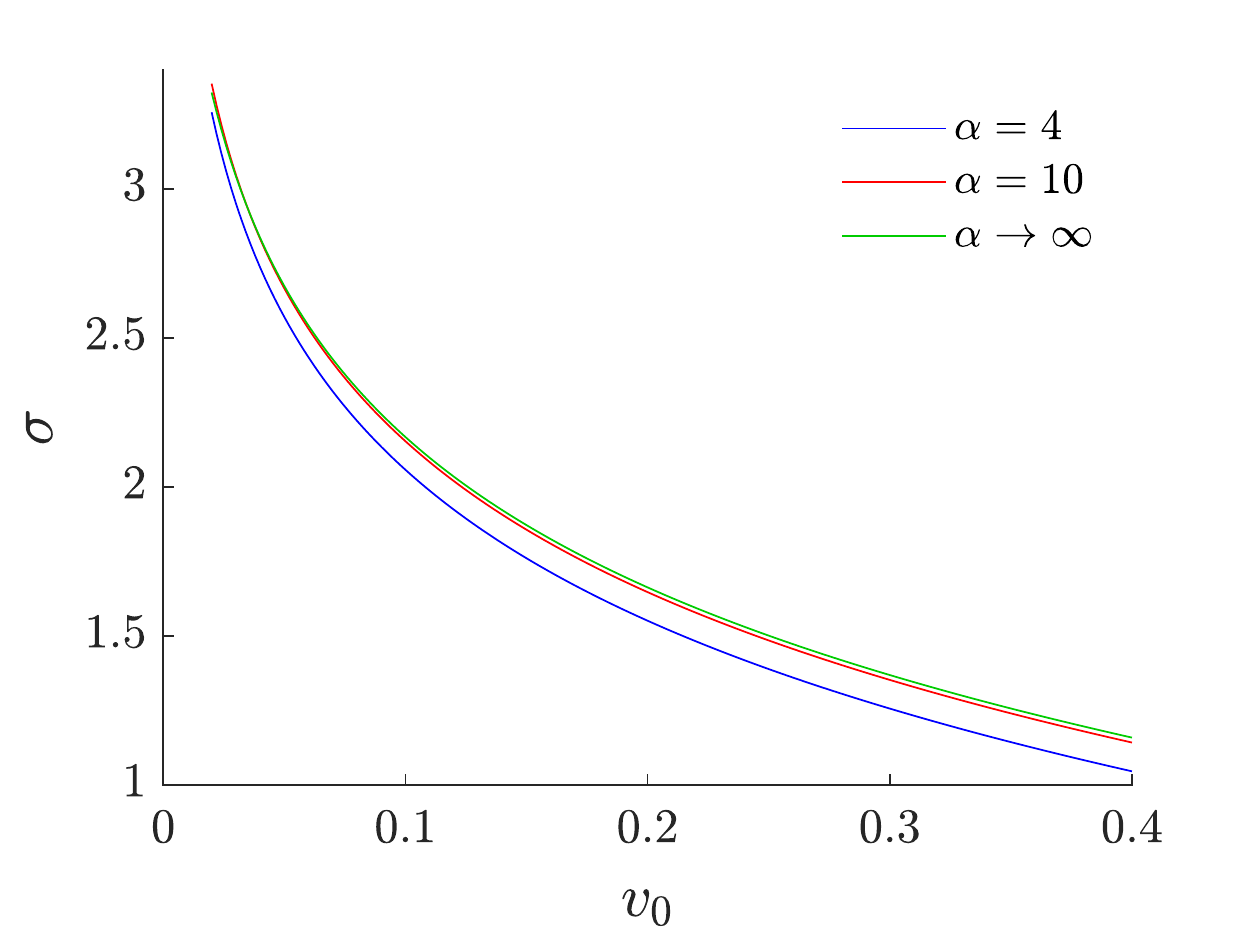} &
\includegraphics[width=9cm]{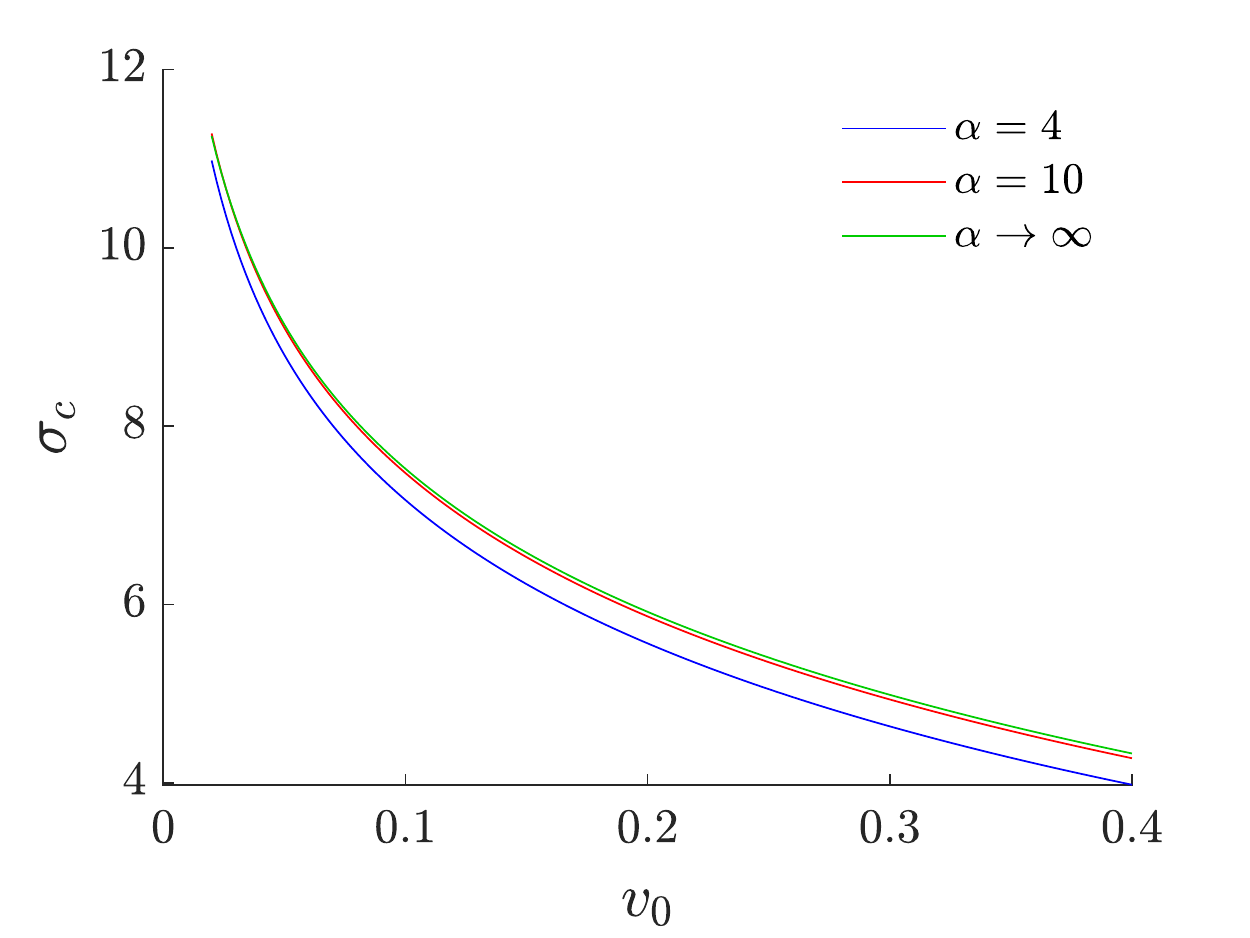} \\
\includegraphics[width=9cm]{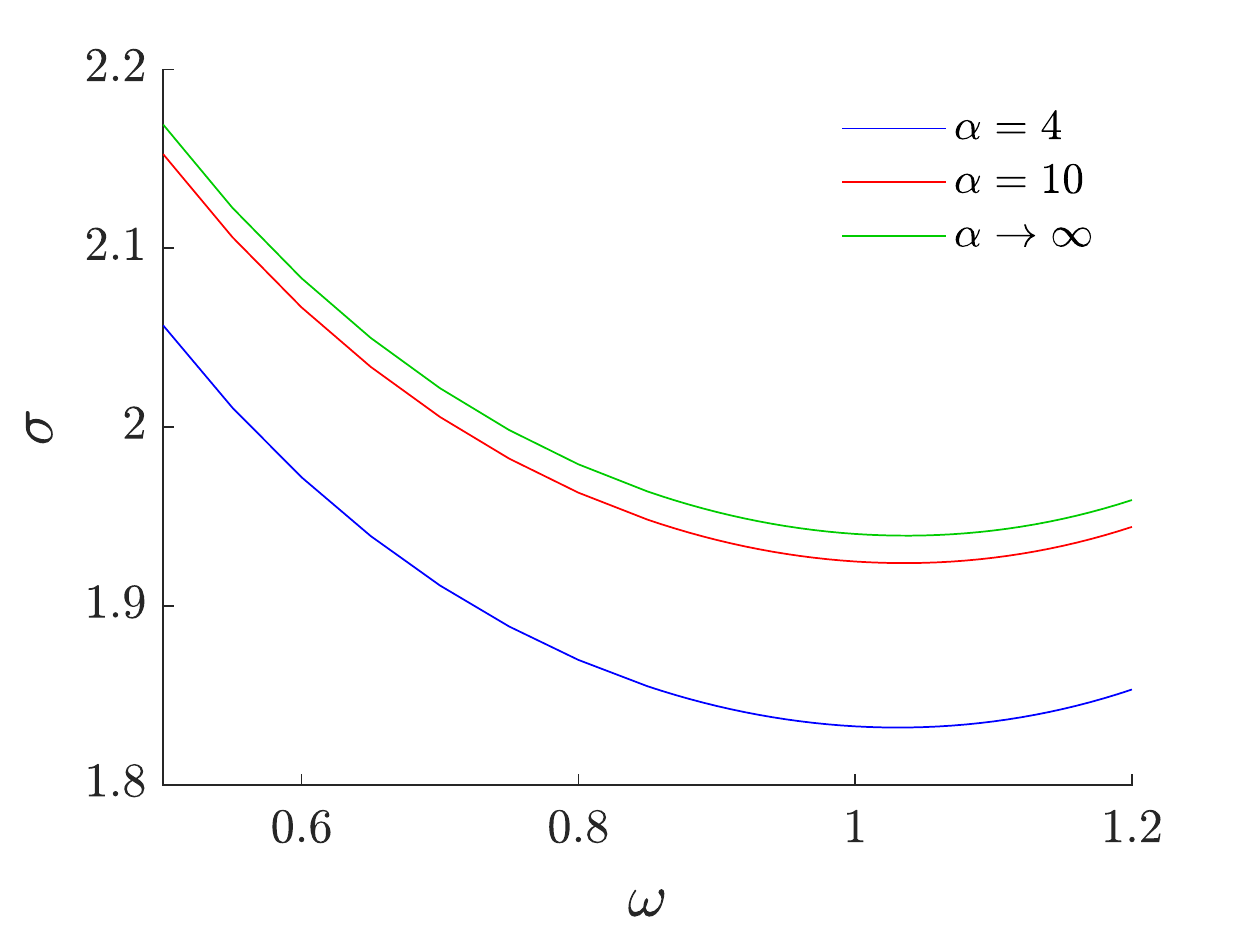} &
\includegraphics[width=9cm]{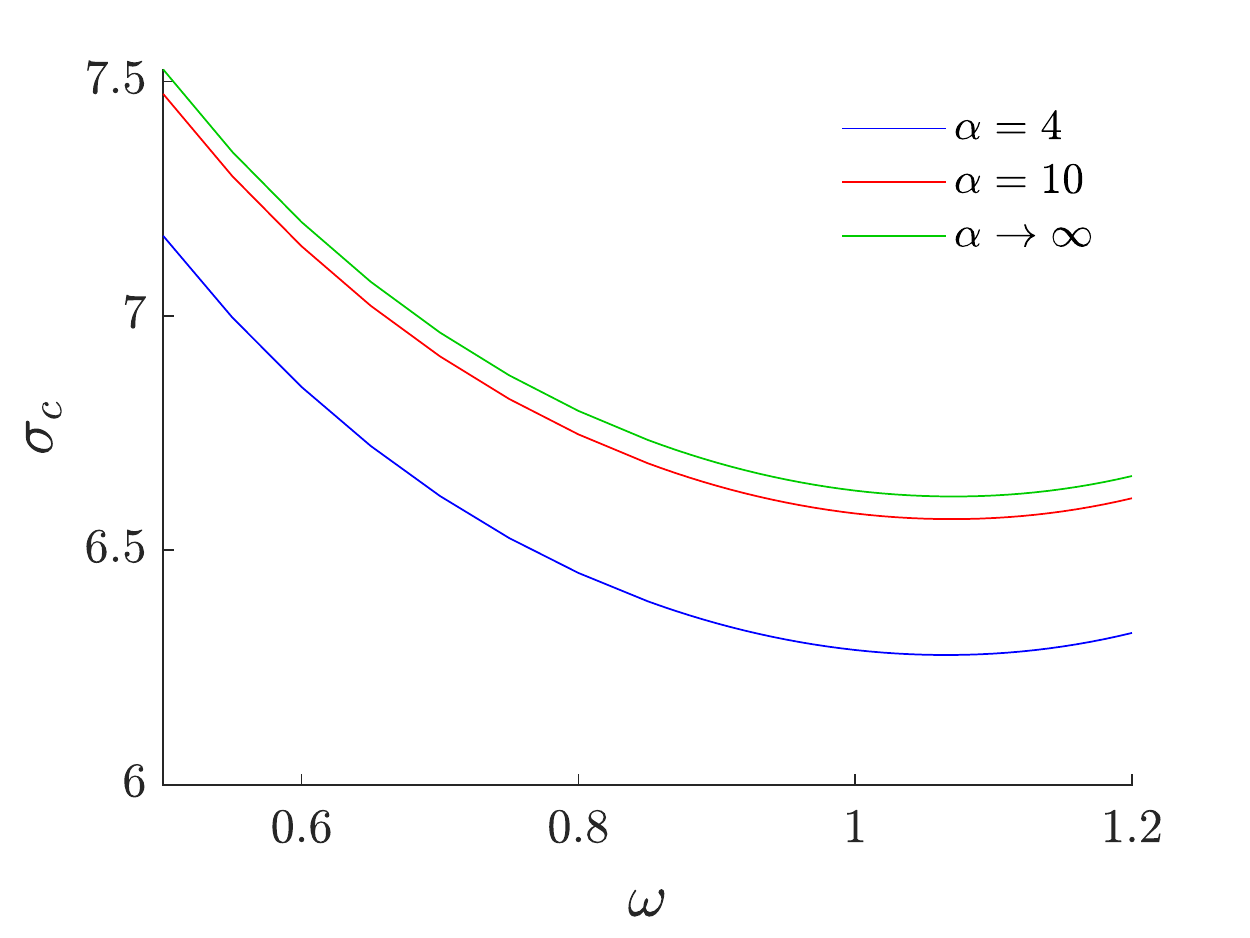} \\
\end{tabular}%
\caption{Dependence of the domain shrinking case by the Goos-H\"anchen effect for $\theta=\cos^{-1}(3/5)$ (left panels) and $\theta=\cos^{-1}(4/5)$ (right panels) for fixed frequency $\omega=0.5$ (top panels) and fixed initial speed $v_0=0.1$ (bottom panels). Dirichlet boundary conditions are denoted by $\alpha\rightarrow\infty$.}
\label{fig:GHE2}
\end{figure}

The above description based on the GHE enables a
systematic understanding of the solitary wave
interaction with a square billiard.
Figure~\ref{fig:GH3} shows the evolution of a classical particle in a shrunk domain by the GHs corresponding to a solitary wave in the presence of a barrier with $\alpha=4$. This evolution is quantitatively compared
with that of the center of mass of the coherent structure. Notice that until the second collision, the two trajectories overlap; the separation between them starts to
subsequently appear because the wave interacts
without following a perfectly straight line, but
rather has a curved part in its trajectory.
This weak effect compounds itself over time, eventually
leading to the non-closure of the orbit.
We have also checked that the relevant situation
does not improve for larger $\alpha$, or for
Dirichlet boundary conditions.
While this procedure allows for a  highly accurate
description of
individual collisions, slight deviations from
straight motion of our deformable effective particles
ultimate lead to the non-closed trajectory,
generic phenomenology
observed herein.

\begin{figure}[tbp]
\begin{tabular}{cc}
\includegraphics[width=9cm]{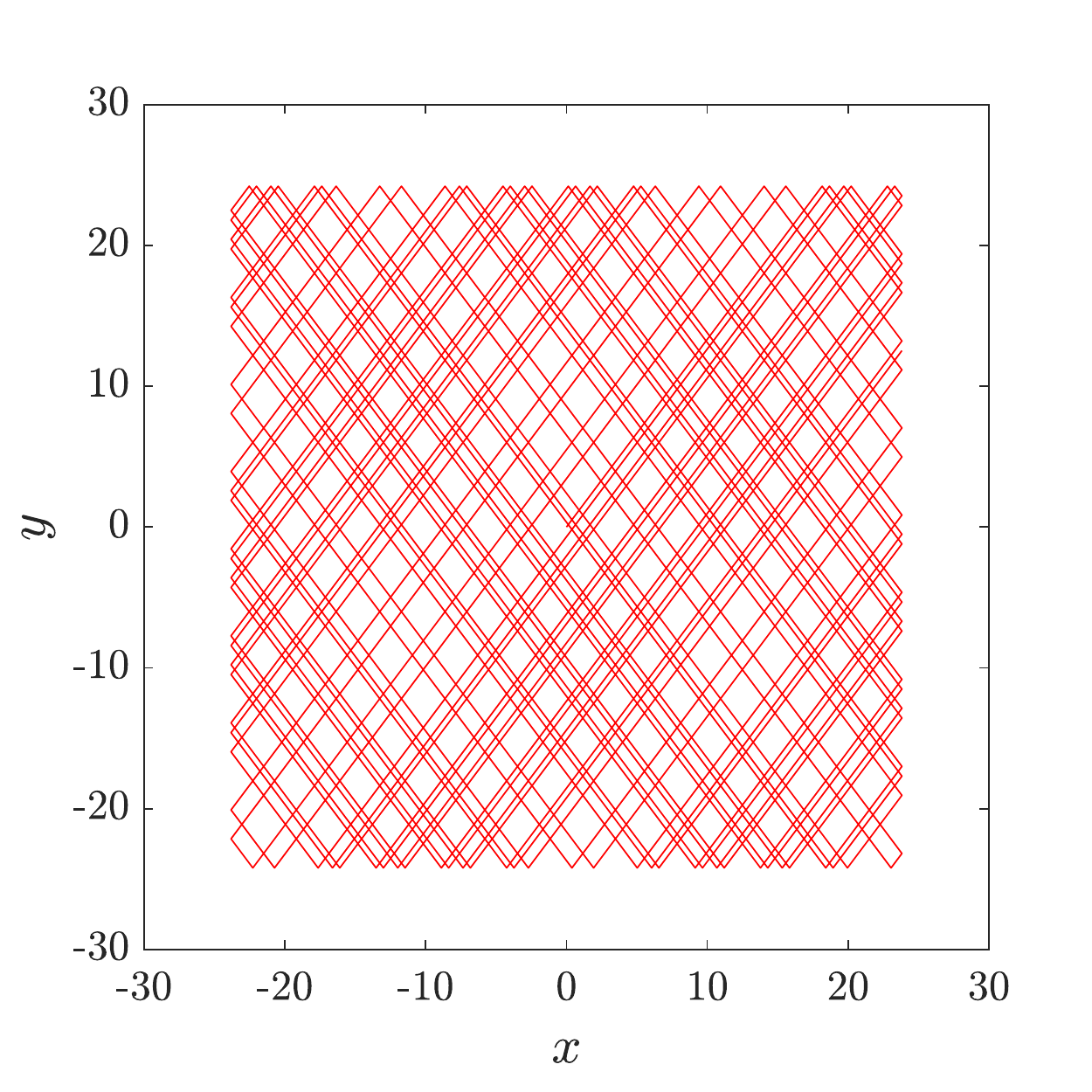} &
\includegraphics[width=9cm]{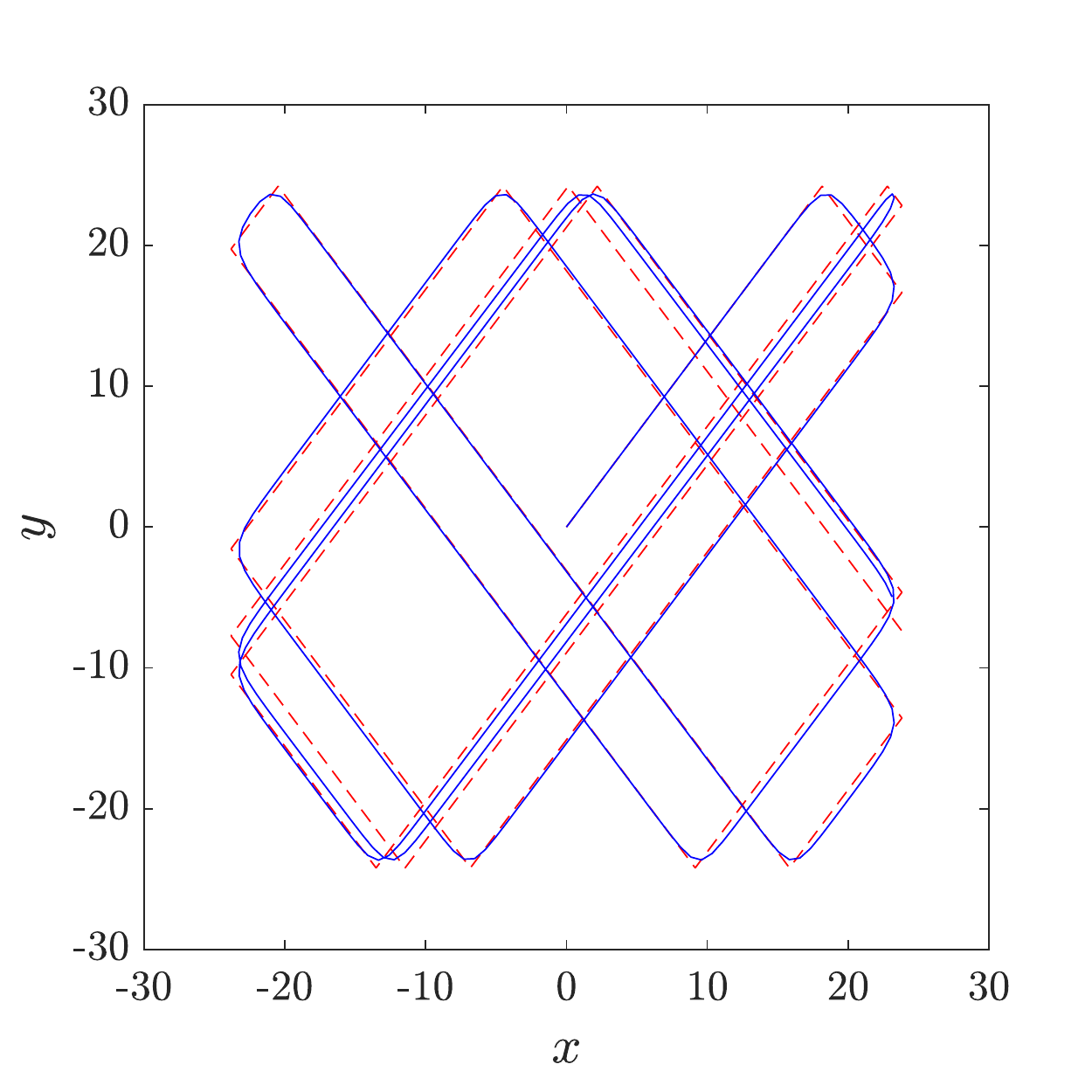} \\
\end{tabular}%
\caption{Trajectory of a classical particle in a shrunk domain with a Goos-H\"anchen shift corresponding to a solitary wave in a square barrier potential of height $\alpha=4$ with the domain shrinking corresponding to $v_0=0.02$ and $\omega=0.5$. The left panel shows the evolution until $t=2\times10^5$, whereas right panel captures the early stage up to $t=34880$ and compares it with the evolution of the the center-of-mass of the solitary wave (full blue line). One can observe that up to the second collision with the barrier, the trajectories overlap, while at later
times, they slightly deviate from each other.}
\label{fig:GH3}
\end{figure}

{The fact that the domain shrinking caused by the GHE is different
  for $\theta$ and $\theta_c$ leads the solitary waves billiard to
  behave effectively as if it is in a rectangular domain. In that
  case, closed orbits in classical billiards can only occur when
  $(a/b)\tan\theta\in\mathbb{Q}$ with $a$ and $b$ being the rectangle
  sides. In our case, if $\tan\theta\in\mathbb{Q}$ and supposing that
  the solitary waves billiard behaves as a classical particle in a
  shrunk domain of sides $3L/2-2\sigma$ and $3L/2-2\sigma_c$, closed
  orbits could only be found in some
  isolated points within the parameter space. However, there is a case for the classical billiard where closed orbits can occur in the shrunk domain, namely this in which $\theta=\theta_c$. This is possible when $\theta=45^\circ$. Because of this, we have launched a soliton with this initial angle, but displaced from the domain center; in particular, we have initialized the soliton from the point $(-3L/8,3L/8)$ which would give rise to a square trajectory in the case of a classical particle. Fig.~\ref{fig:45degrees} shows the trajectory for a soliton launched with an initial velocity $v_0=0.1$. This trajectory is not a square because the initial point is not at the center of the second quadrant of the shrunk domain. Notice that the closed trajectory can be found for $v_0\lesssim0.25$, as for larger values the radiation caused by the soliton impacting on the barrier prevents from the trajectory closure.}

\begin{figure}[tbp]
\begin{tabular}{cc}
\includegraphics[width=9cm]{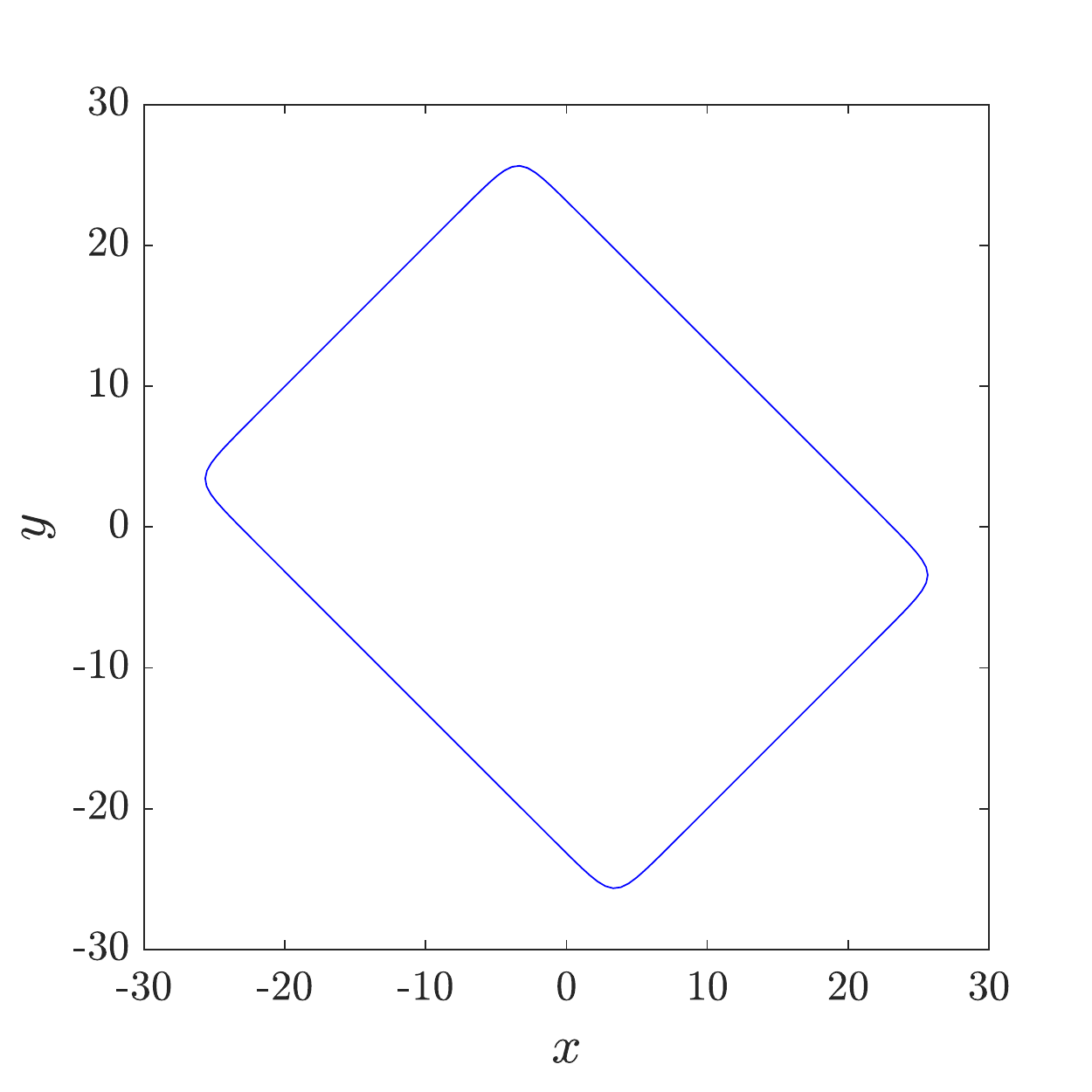} &
\includegraphics[width=9cm]{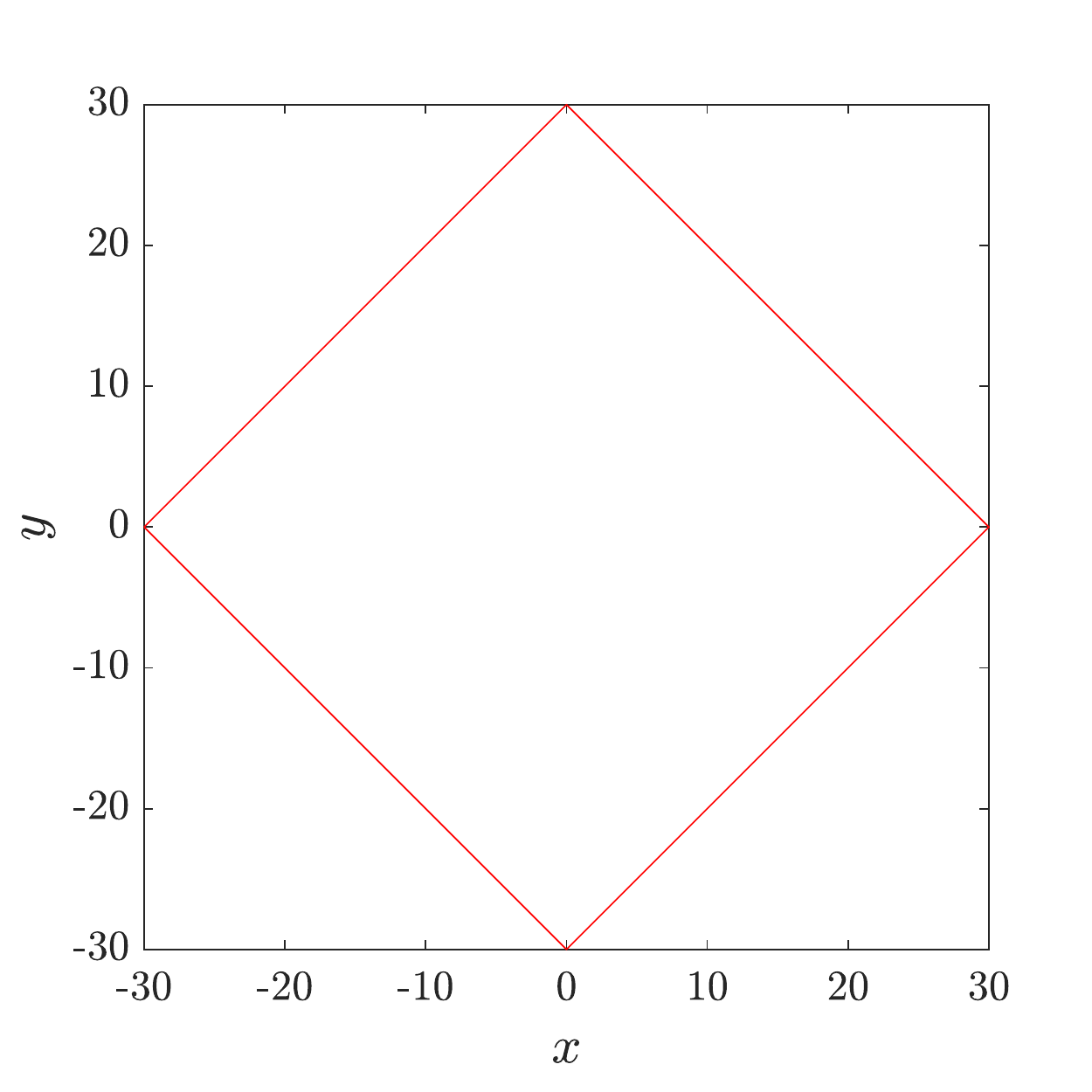} \\
\end{tabular}%
\caption{(Left) Evolution of the center-of-mass of the finite-width solitary wave with $\alpha=4$ and $v_0=0.1$, for $\omega=0.5$ launched from $(-15,15)$ with an initial angle $\theta=45^\circ$. See companion movie \texttt{movie5.gif} in \cite{movies}, in which the launching point is marked by a white dot in order to observe that the soliton comes back to that point. (Right) Trajectory of a classical {\it point} particle launched from the same site and initial angle.}
\label{fig:45degrees}
\end{figure}

\subsection{Some other prototypical solitary wave billiards}

We end this section by showing the outcome of the solitary
wave dynamics in a Bunimovich stadium and a Sinai billiard (see the left top and bottom panels, respectively, of Fig.~\ref{fig:stadium}).
Here, we are motivated by the chaotic nature of these
billiards even for classical point particles and
we seek to observe the dynamics of our deformable
effective particle therein.
In the first case, the barrier (composed by a square whose sides have a length 48 attached to two semicircles of radius 28) has a potential height of $\alpha=4$ and is embedded into a domain $[-64.5,64.5]\times[-32.25,32.25]$, and the solitary wave is launched from $(0,0)$. In the second case, the domain size is $[-75,75]\times[-75,75]$, encompassing a square barrier of side length 112.5 and a circular barrier of radius 2, both centered at $(0,0)$, that have a potential height of $\alpha=4$. Here, the solitary wave is launched from the
point $(-20,-20)$. The path followed by the center of mass of the wave of initial velocity $v_0=0.02$ is depicted in the right panels of Fig.~\ref{fig:stadium}. In both cases, we can observe the chaotic nature
of the dynamics which is reminiscent of the ergodic nature of the
point particle case, yet again with a significant difference borne
out of the GHE present in the solitonic case. The latter effect
naturally, in this case as well, precludes the accessibility
of a substantial region of the configuration space,
similarly to the above presented square billiard case.

\begin{figure}[tbp]
\begin{tabular}{cc}
\includegraphics[width=9cm]{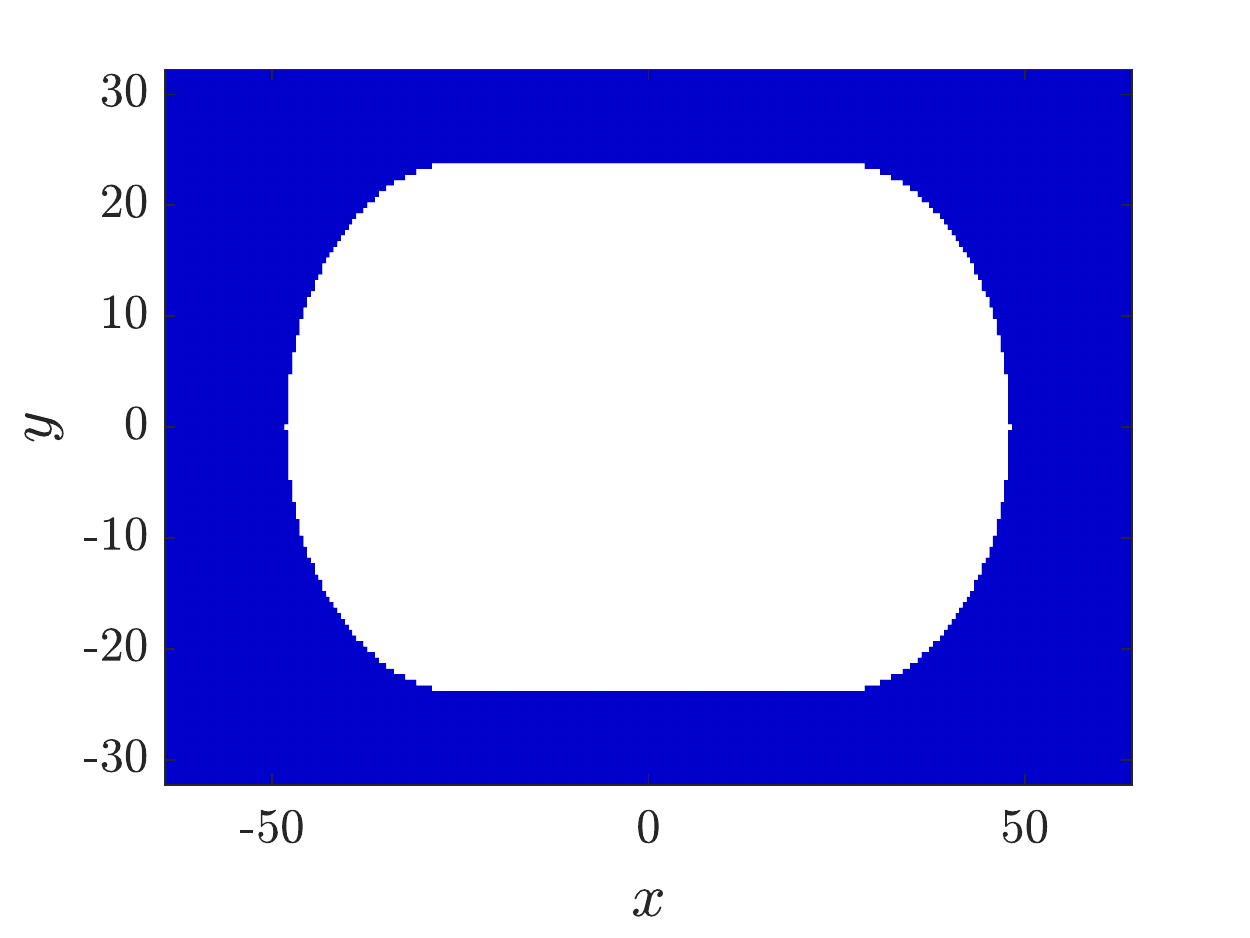} &
\includegraphics[width=9cm]{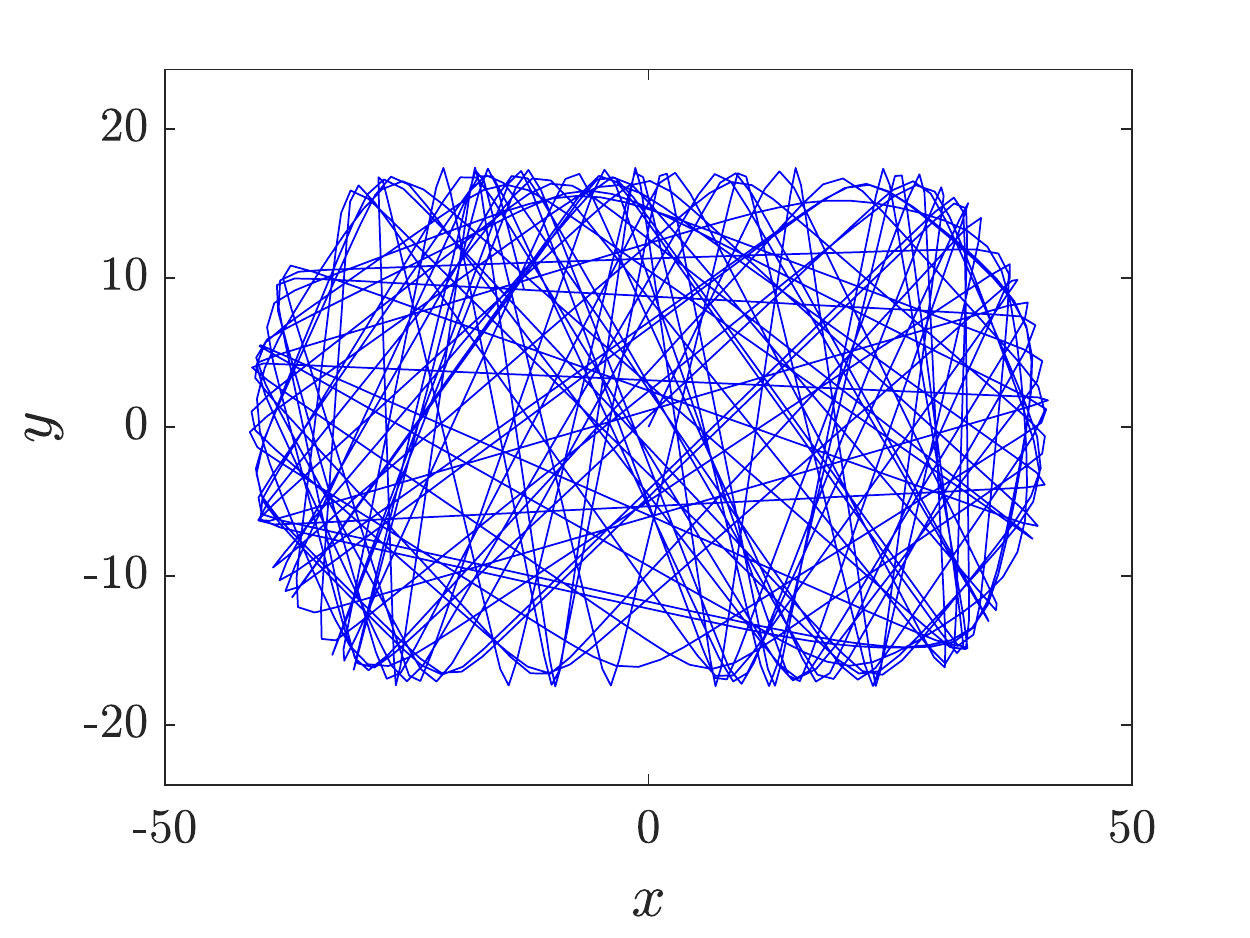} \\
\includegraphics[width=9cm]{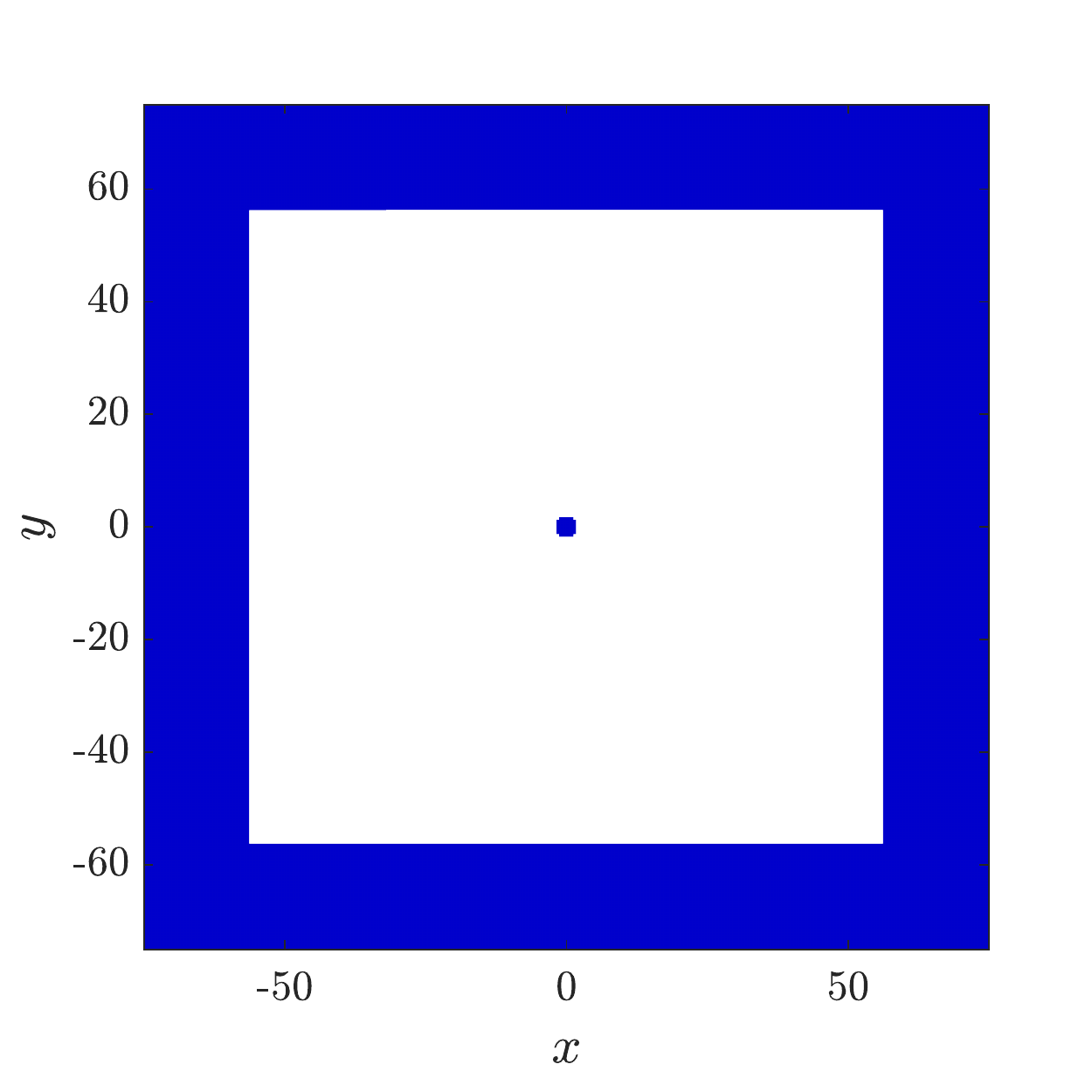} &
\includegraphics[width=9cm]{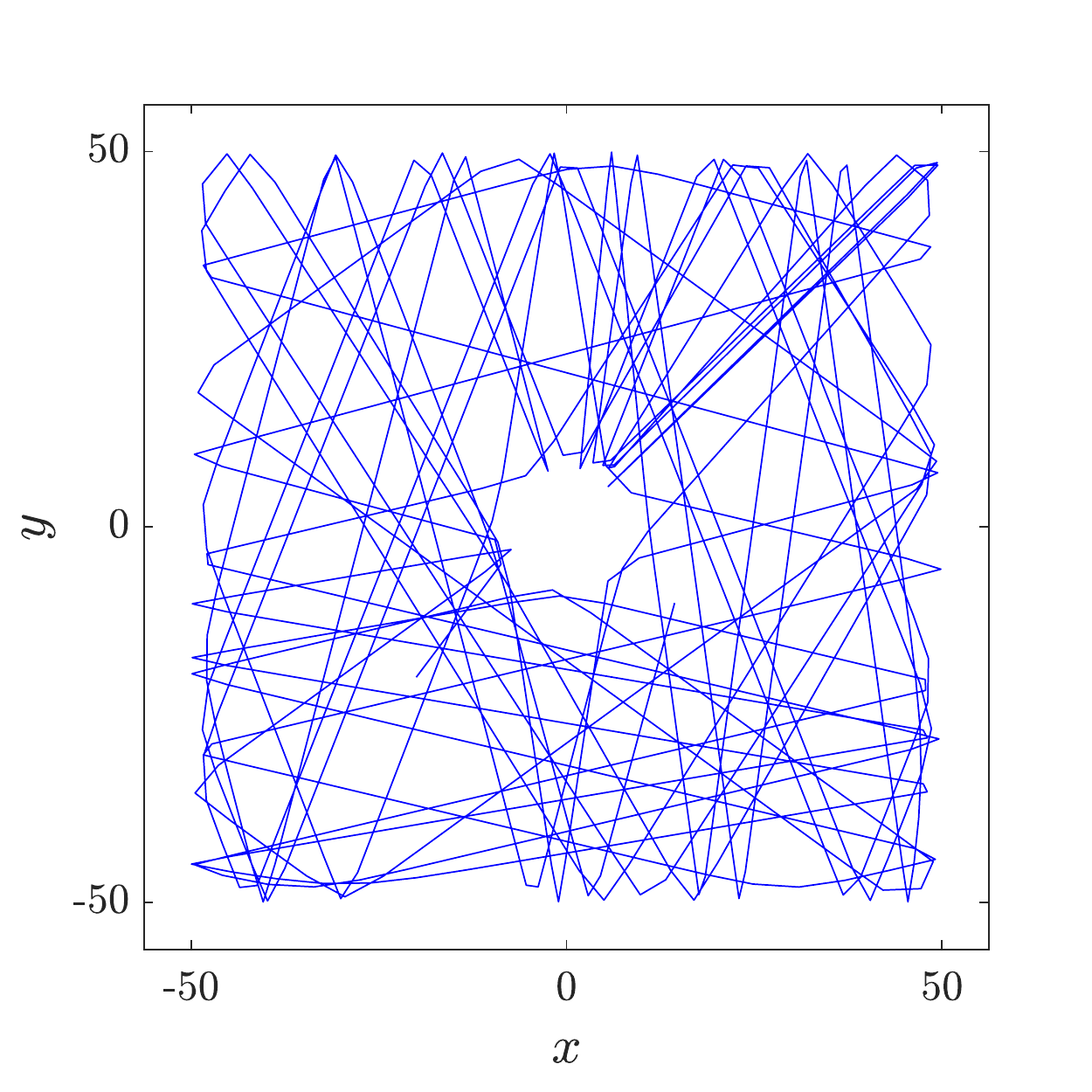} \\
\end{tabular}%
\caption{Left panels: form of the Bunimovich stadium and Sinai billiard potentials.
Right panels: evolution of the center-of-mass of the solitary wave with $v_0=0.02$ in a Bunimovich stadium (left panel) and a Sinai billiard (right panel). See also companion movies \texttt{movie6.gif} and \texttt{movie7.gif} in \cite{movies} depicting the evolution of the billiard dynamics. The final time of both simulations is $t=3.2\times10^5$.}
\label{fig:stadium}
\end{figure}

\section{Conclusions and Future Challenges}

In the present work, we have proposed a new paradigm for the dynamics
of solitary waves, namely the study of solitonic billiards.
We have judiciously selected a model that, on the one hand,
is physically relevant, yet on the other hand avoids well-known
pathologies of higher-dimensional NLS models, such as the presence
of self-focusing and wave collapse~\cite{sulem}. By choosing the
saturable nonlinearity model, we present a setting of relevance, e.g.,
to photorefractive optical crystals that possesses a Hamiltonian
nature (in line with classical point-particle billiards), but which
also has other key advantages, such as the ability to Galilean-boost
the coherent structures.
{ For this model, we identify the full
branch of fundamental solitary wave solutions
as the frequency of the standing waves is varied.}
We identified some important differences that this solitonic
billiard features in comparison to the point particle case.
Indeed, the finite width of the solitary particle and the
non-integrable nature of the model present
new possibilities including that of the collision storing some
of the kinetic energy into internal mode oscillations, as well
as that of the collision not being perfectly elastic.
{ Both of these features have been
observed in direct numerical simulations
with sufficiently high speeds or/and
involving sufficiently wide solitary waves.}
At the same time, we have observed the presence of a solitonic
analogue of the (negative) Goos-H{\"a}nchen effect which also leads to significant
deviations from the point particle case, including
the effective shrinkage of the billiard domain. These features combined
render even integrable point particle billiards non-integrable
ones when considered in this solitonic realm (such as, e.g., the
square/rectangular billiards).
{ Nevertheless, our simulations have
demonstrated that a reduced-domain approach
accounting for the GHE can adequately follow the
solitary wave scattering for a long time. For
collisions with incident angles of $\theta=45^\circ$,
it was shown that the trajectories can be
closed for sufficiently low speeds.}
For completeness, we also
examined the case of billiards such as the Bunimovich
stadium or the Sinai billiard that are chaotic at the point-particle
level and observed similar phenomenology (but with the above features
still manifest) in the present setting.

We believe that this vein of studies is at a nascent stage
and hence there are numerous opportunities for the future.
It is interesting, for instance, to explore how the variation of
the solitonic width (controlled by $\omega$) would change the
case considered herein.
Furthermore, it would be of relevance to consider
the Lyapunov exponents of the present
wave billiard dynamics and to compare it with the
corresponding quantities for the effective particle
system, as a vein of quantitative comparison of the
two systems, across different billiard settings.
Such an understanding could lead to leverage the effective
particle model towards analytical calculations in the future~\cite{valagia}.
Another possibility would be to examine
some case examples of models that are integrable in
the two-dimensional setting such as the Davey-Stewartson (or
the Kadomtsev-Petviashvili) equation and explore how their integrable
structures may fare in terms of similarities and differences
to the non-integrable ones considered herein. It would also
be of relevance to examine corresponding generalizations to higher
dimensional settings and explore billiard enclosures and solitonic
motion therein.

{ Yet another dimension of possible experimental
  realizations
  can be considered in the setting of atomic condensates, following
  the earlier experimental work of~\cite{raizen}.
  There, given
the cubic nature of the predominant nonlinearity, the attractive or self-focusing case
would not be immediately accessible~\cite{pethick,stringari} due to
the
potential of this nonlinearity for self-similar
colapse~\cite{sulem}. However,
it would be eminently relevant to
examine the self-defocusing or self-repulsive setting in 2d where the prototypical
patterns would be topologically charged vortices and to explore their
dynamics ~\cite{siambook} in enclosures such as the ones considered
herein. For a related recent example of BEC in a optical box trap,
see, e.g.,~\cite{box}.
Extensions thereof that may bear more of a ``Newtonian particle'' character due to a bright solitary wave in a second component have also been recently
analyzed; see, e.g.,~\cite{richaud} for a relevant example.}
These are only some of the possible directions
emerging. Some of these are currently under consideration and
will, hopefully, be reported upon in future publications.

\vspace{10mm}

{\it Acknowledgments.}
This material
is based upon work supported by the US National
Science Foundation under Grants DMS-1809074, PHY-
2110030 (P.G.K.). P.G.K. also gratefully
acknowledges a query of Prof. Tsampikos
Kottos that initiated this vein of research and a discussion with
Prof. Mason Porter. J.C.-M. acknowledges support from EU (FEDER program 2014-2020) through both Consejería de Economía, Conocimiento, Empresas y Universidad de la Junta de Andalucía (under the projects P18-RT-3480 and US-1380977), and MCIN/AEI/10.13039/501100011033 (under the projects PID2019-110430GB-C21 and PID2020-112620GB-I00).

\vspace{10mm}

\end{document}